\documentclass[12pt]{article}

\usepackage{amsmath,euscript,array,amssymb,cite,mathrsfs,theorem,fancyhdr,booktabs} % use drftcite in draftmode otherwise
\usepackage{epsfig}

{\theorembodyfont{\rmfamily}
\newtheorem{note}{}}
\numberwithin{note}{section}

\numberwithin{equation}{section}
\newcommand{\BOX}[1]{
\begin{center}\fbox{\parbox{15cm}{#1}}\end{center}}

\def\N{{\EuScript N}}
\def\bN{{\boldsymbol N}}

\newcommand{\MS}{$\overline{\text{MS}}$}

\newcommand{\MAT}[1]{\begin{pmatrix}#1\end{pmatrix}}

\newcommand{\EQ}[1]{\begin{equation}\begin{split} #1 \end{split}\end{equation}}

\newcommand{\SP}[1]{\begin{equation}\begin{split} #1
    \end{split}\end{equation}}
\newcommand{\EQN}[1]{\begin{equation} #1 \end{equation}}
\newcommand{\SPN}[1]{\begin{equation}
\begin{split} #1 \end{split}\end{equation}}

\def\aD{{\dot\alpha}}

\def\LAG{\mathscr{L}}

\def\PF{\mathscr{Z}}
\def\FF{\mathscr{F}}
\begin{document}

\addtolength{\baselineskip}{4pt}
\thispagestyle{empty}

%\begin{flushright}
%***\\
%{\tt hep-th/0206063}\\
%May 2002
%\end{flushright}

\phantom{.}

\vspace{2cm}

\begin{center}
{\scshape\Huge 6 Lectures on QFT, RG and SUSY} 

\vspace{1.2cm}

{\scshape Timothy J.~Hollowood}

\vspace{0.15cm}

{\sl Department of Physics, Swansea University,\\
Swansea, SA2 8PP, UK}\\ {\tt t.hollowood@swansea.ac.uk}\\

\vspace{2cm}
Lectures given at the British
Universities Summer School in Theoretical Elementary Particle
Physics (BUSSTEPP) Liverpool University August 2009 and
Cambridge University August 2008.\\

\vspace{1cm}
{\bf\Large Abstract}
\end{center}

\vspace{0.5cm}
An introduction to the theory of the renormalization group in
the context of quantum field theories of relevance to particle physics
is presented in the form of 6 lectures delivered to the British
Universities Summer School in Theoretical Elementary Particle
Physics (BUSSTEPP). Emphasis is placed on gaining a physical understand of the
running of the couplings and the Wilsonian version of the
renormalization group is related to
conventional perturbative calculations with dimensional regularization
and minimal subtraction. 
An introduction is given to some of the remarkable
renormalization group properties of supersymmetric theories.

\newpage

%\section*{Contents}
%\def\MS{\phantom{a}}

%\contentsline {section}{\numberline {I}\hspace{0.5cm}Introduction}{7}
%\contentsline {subsection}{\numberline {I.1}\hspace{0.5cm}The philosophy of instanton calculations}{11}
%\contentsline {section}{\numberline {II}\hspace{0.5cm}Instantons in Pure Gauge Theory}{13}
%\contentsline {subsection}{\numberline {II.1}\hspace{0.5cm}Some basic facts}{13}
%\contentsline {subsection}{\numberline {II.2}\hspace{0.5cm}Collective coordinates and moduli space}{15}
%\contentsline {subsection}{\numberline {II.3}\hspace{0.5cm}General properties of the moduli space of instantons}{19}
%\contentsline {subsubsection}{\numberline {II.3.1}\hspace{0.5cm}The moduli space as a complex manifold}{19}

%\tableofcontents

The purpose of these lectures is to introduce a powerful way to think
about Quantum Field Theories (QFTs). 
This conceptual framework is Wilson's version of 
the Renormalization Group (RG).
The only pre-requisites are a basic understanding of QFTs along the lines
of a standard introductory QFT course: the Lagrangian formalism,
propagator, Feynman rules, the path integral formulation, {\it etc\/}.

It turns out that supersymmetric theories are a wonderful arena for
discussing RG because there are many things that one can prove
exactly. For this reason the last two lectures will provide a very
basic description of some of the extraordinary features that SUSY
theories have in regard to the RG. The discussion of SUSY will
necessarily be very rudimentary.

I have not included any references in these notes. However, sources
which I have found particularly useful are:

(1) The idea of the renormalization group goes back quite a long way. 
In this course we have in mind the version of the renormalization group
    due to Wilson and one can do no better than consult the Physics
    Reports (Volume 12, Number 2, 1974) by Wilson and Kogut.

(2) Peskin and Schroeder's textbook on QFT ``Quantum Field Theory'' is
    probably the most useful and will fill in many of the gaps I have
    left. I also like Zinn-Justin's huge
    ``Quantum Field Theory and Critical Phenomena''.

(3) Weinberg's notes on ``Critical Phenomena for Field Theorists'' are
very useful for explaining how many of the ideas of the
renormalization group came from statistical physics and are 
available in scanned form
at {\tt http://ccdb4fs.kek.jp/cgi-bin/img/allpdf?197610218}.

(4) Manohar's notes on ``Effective Field Theories'' {\tt
    hep-ph/9606222} have an excellent discussion of the delicate issue
    of decoupling in momentum-dependent and -independent RG schemes.

(5) Strassler's ``An Unorthodox Introduction to Supersymmetric Gauge
    Theory'' are not only unorthodox but excellent because they tackle
        many issues that are not covered in conventional SUSY texts.

(6) I will not use superspace when discussing SUSY, but a standard
    source for this booking-keeping device 
is the book ``Supersymmetry and Supergravity'' by Wess and Bagger.

(7) Our discussion of the beta function of SUSY gauge theories has
    been heavily influenced by 2 very insightful
  papers by Arkani-Hamed and Murayama, {\tt hep-th/9705189} and {\tt
  hep-th/9707133} which clarify (at least for me) many of the
  confusing issues regarding renormalization in SUSY gauge theories.

Naturally, a series of only 6 lectures can only scratch the surface of
this subject and I have had to leave out lots of things and simply the
discussion. Some
additional information appears in the form of a set of notes at the
end of each lecture.

I would like to thank the organizers of BUSSTEPP, Jonathan Evans in
Cambridge 2008 and Ian Jack in Liverpool 2009, for providing 
excellently run summer schools that enabled me to develop my idea to
teach QFT with the renormalization group as the central pillar.

\newpage
\section{The Concept of the Renormalization Group}

The key idea of the renormalization group results from comparing phenomena 
at different length/energy scales. A QFT is defined by an
action functional of the fields $S[\phi;g_i]$ which depends {\it a
  priori\/} on an
infinite number of parameters: the coupling constants in a general
sense so including mass parameters, {\it etc.\/}. 
The set of couplings $\{g_i\}$ can be thought of
as a set of coordinates on
{\it theory space\/}. The functional integral takes the form
\EQN{
\PF=\int [d\phi]\,e^{iS[\phi;g_i]}
}
and so, as well as the action, we have to define the measure
$\int[d\phi]$. This is a very tricky issue since a classical field has
an infinite number of degrees-of-freedom and it is by no means a
trivial matter to integrate over such an infinite set of variables. 
In perturbation theory a symptom of the difficulties in defining the
measure shows up as the divergences that occur in loop integrals. These UV
divergences occur when the momenta on internal lines become large so
they are intimately bound up with the fact that the field
has an infinite number of degrees-of-freedom and can fluctuate on all
energy scales. Of course there may also be IR divergences, however these are
not as conceptually serious as the UV ones: in reality in a real
experiment one is working in some finite region of spacetime and this
provides a natural IR cut-off.

At least initially, in order to make sense of $\int[d\phi]$, we have
to implement 
some UV cut-off procedure in order to properly define the measure,
or equivalently, in perturbation theory 
regulate the infinities that occur
in loop diagrams. As we have said above, these UV, high energy
divergences occur because the fields can fluctuate at arbitrarily
small distances and in order to regulate the theory we have to somehow
suppress these high energy modes. Whatever way this is done
inevitably introduces a new 
energy scale $\mu$, the {\it cut-off\/}, into the theory. 

\BOX{\begin{center}{\bf Cut-offs or Regulators}\end{center}

There are many ways of introducing a cut-off, or regulator, into a
  QFT. For example, one can define the theory on a spatial lattice
  (after Wick rotation to Euclidean space). In this case $\mu^{-1}$ is
  the physical lattice spacing. Or one can suppress the high
  momentum modes by modifying the action or the measure. Or in
  perturbation theory one can analytically continue the spacetime
  dimension, a procedure known as dimensional regularization.}

Suppose we have some
physical quantity $\FF(g_i;\ell)_\mu$ 
which  can depend in general on a (or possibly several) 
 length scale $\ell$ (or equivalently an energy scale $1/\ell$). 
The theory of RG postulates that one can change the cut-off 
of the theory in such a way that the physics on energy
scales $<\mu$ remains constant. In order that this is possible the 
couplings must change with $\mu$. This idea can be summed up in the
RG equation:$^{\ref{n11}}$
\EQ{
\FF(g_i(\mu);\ell)_\mu=\FF(g_i(\mu');\ell)_{\mu'}\ .
\label{haq1}
}
The functions $g_i(\mu)$ with
defines the RG flow of the theory in the space of
couplings. The RG flow is
conventionally thought of as being 
towards the IR, {\it i.e.\/}~decreasing $\mu$, but we shall often think
about it in the other direction as well, towards the UV with $\mu$ increasing.
In order that the RG equation \eqref{haq1} can hold it is necessary
that the space of couplings includes {\it all\/} possible couplings
(necessarily an infinite number).
The RG is non-trivial because in order to lower the cut-off we somehow
have to ``integrate out'' the degrees-of-freedom of the theory that lie
between energy scales $\mu$ and $\mu'$. In general this is a 
difficult step, however, as we shall see, in QFT we are
in a very lucky situation due to the remarkable focusing 
properties of RG flows.

The RG energy scale $\mu$ plays a central r\^ole in the theory and it
is important to understand what exactly it is. To start with we have
identifed $\mu$ with the physical cut-off; however, there is another
way to interpret $\mu$. The point is that if we wish to describe
physical process at the energy scales $E_\text{phys}$ or below, 
or distance scales greater than
$\ell=E^{-1}_\text{phys}$, 
then there is no reason why we cannot take the cut-off
to be at the scale $\mu=E_\text{phys}$. In fact this would be the
optimal choice since the efffective description would then only
involve modes with energies $\leq E_\text{phys}$, {\it i.e.\/}~the
ones directly involved in the physical process. So another to think of
RG flow is that the couplings run with the typical energy scale of the
process being investigated, $g_i(\mu)$, where $\mu$ is that energy scale.

Since the physical observables of the theory can be determined once
the action is known, the RG transformation itself follows from
following how the action changes as the cut-off changes. The action at
a particular cut-off is
known as the Wilsonian Effective Action $S[\phi;\mu,g_i]$ and since it
depends on the fields, as well as the couplings, 
the RG transformation must be generalized to
\BOX{\begin{center}{\bf The Key RG Equation}\end{center}
\EQ{
S\big[Z(\mu)^{1/2}\phi;\mu,g_i(\mu)\big]=
S\big[Z(\mu')^{1/2}\phi;\mu',g_i(\mu')\big]\ ,
\label{haq3}
}}
where $Z(\mu)$ is known as 
``wavefunction renormalization'' of the field.
In the general case with many fields, $Z(\mu)$
is a matrix quantity that can mix all the fields. 
The action of a QFT can be written as the sum of a kinetic term and 
linear combination of ``operators'' 
${\cal O}_i(x)$ which are powers of
the fields and their derivatives, {\it e.g.\/} $\phi^n$, $\phi^n
\partial_\mu\phi\partial^\mu\phi$, {\it etc\/}:$^{\ref{n12}}$
\EQ{
S[\phi;\mu,g_i]=\int d^dx\,\Big[\frac12\partial_\mu\phi\partial^\mu
\phi+\sum_i \mu^{d-d_i}g_i{\cal O}_i(x)\Big]
}
where $d_i$ is the classical dimension of ${\cal O}_i(x)$. Notice that
we chosen the couplings to be dimensionless by inserting the
appropriate power of the cut-off to soak up the dimensions. This is
because it is really the value of a coupling relative to the cut-off
that is physically relevant. The wavefunction renormalization factor
$Z(\mu)$ can be thought of as the coupling to the kinetic term
since$^{\ref{n112}}$  
\EQ{
S[Z^{1/2}\phi;\mu,g_i]=\int d^dx\,\Big[\frac Z2\partial_\mu\phi\partial^\mu
\phi+\cdots\Big]
\ .
}

It is often useful to think about infinitesimal RG transformations, in
which case we define the {\it beta function\/} of a theory
\EQ{
\mu\frac{dg_i(\mu)}{d\mu}\ .
\label{bfun}
}
The running couplings then follow by integration of the beta-function
equations above.
Notice that due to the fact that couplings always
    appear as combinations $\mu^{d-d_i}g_i$ means that 
the beta functions always have the form
\EQN{
\mu\frac{dg_i}{d\mu}=(d_i-d)g_i+
\beta^\text{quant.}_{g_i}\ .
}
One can think of the first term as arising from classical scaling
and the second piece as arising 
from the non-trivial integrating-out part of the RG transformation.
If $\hbar$ were re-introduced $\exp[iS]\to\exp[iS/\hbar]$, then the 
quantum piece would indeed vanish in the limit $\hbar\to0$.
We can also define the 
{\it anomalous dimension\/} of a field $\phi$ as 
\EQN{
\gamma_\phi=-\frac \mu2\frac{d\log Z(\mu)}{d\mu}\ .
}

In particle physics the ultimate physical observables are the
probabilities for particular things to happen.
However, it is often useful, especially in massless theories
where the S-matrix is problematic to formulate because of long-range
interactions, to consider the Green
functions of fields. Schematically,
\EQN{
\big\langle\phi(x_1)\cdots\phi_n(x_n)\big\rangle_{g_i(\mu),\mu}
=\frac{\int_\mu [d\phi]\,
e^{iS[\phi;\mu,g_i(\mu)]}
\phi(x_1)\cdots\phi(x_n)}{\int_\mu [d\phi]\,
e^{iS[\phi;\mu,g_i(\mu)]}}\ .
}
It follows from \eqref{haq3} that for these quantities that depend on
the field we must generalize \eqref{haq1} to take account of wavefunction 
renormalization, giving 
\EQ{
Z(\mu)^{-n/2}\big\langle\phi(x_1)\cdots\phi_n(x_n)\big\rangle_{g_i(\mu),\mu}
=Z(\mu')^{-n/2}
\big\langle\phi(x_1)\cdots\phi_n(x_n)\big\rangle_{g_i(\mu'),\mu'}\ ,
\label{haq2}
}

What is particularly important about RG flows are their IR and UV
limits; namely $\mu\to0$ and $\mu\to\infty$, respectively. 
As we flow towards the IR, all masses relative to the cut-off, that is 
$m/\mu$, increase. If a theory has a mass-gap
(no massless particles) then as $\mu\to0$ 
all physical masses are become infinitely heavy relative to the cut-off
and there is nothing left to propagate in the IR. Hence, in the IR
limit we have an empty, trivial or null theory. The other possibility is when
the RG flow starts on the:

\BOX{\begin{center}{\bf Critical Surface}\end{center} 

The infinite dimensional subspace in the 
space-of-theories for which the mass gap vanishes. These theories
consequently have a non-trivial IR limit in which only the massless
degrees-of-freedom remain.
}

In this case, as $\mu\to0$ the massless particles
will remain and in all known cases the couplings flow to a fixed
point of the RG $g_i(\mu)\to g_i^*$ as $\mu\to0$ 
where the beta functions vanish:$^{\ref{n15}}$

\BOX{\begin{center}{\bf Equation for a fixed point or conformal field
    theory}\end{center} 
\EQ{
\mu\frac{dg_i}{d\mu}\Big|_{g_j^*}=0\ .
}}

The theories at the fixed points are very special because as well only
having massless states particles they have 
no dimension-full parameters at all. 
This means that they are scale
invariant. However, this scale invariance is naturally promoted to the
group of conformal transformations
and so the fixed point theories are also ``conformal field theories'' 
(CFTs).$^{\ref{n17}}$  

In the neighbourhood of a fixed point, or CFT, $g_i=g^*_i+\delta g_i$, 
we can alway linearize the RG flows:
\EQN{
\mu\frac{dg_i}{d\mu}\Big|_{g_j^*+\delta g_j}=A_{ij}\delta g_j+{\cal
  O}(\delta g_j^2)\ .
}
In a suitable diagonal basis for $\{\delta g_i\}$ which we denote
$\{\sigma_i\}$, 
\EQN{
\mu\frac{d\sigma_i}{d\mu}=(\Delta_i-d)\sigma_i+{\cal O}(\sigma^2)
}
and so to linear order the RG flow is simply
\EQ{
\sigma_i(\mu)=\left(\frac{\mu}{\mu'}\right)
^{\Delta_i-d}\sigma_i(\mu')\ .
\label{RGf}
}
The quantity $\Delta_i$
is called the scaling (or conformal) 
dimension of the operator associated to $\sigma_i$.
In general in an interacting QFT it will not
be the classical scaling dimension and the difference
\EQ{
\gamma_i=\Delta_i-d_i
}
is known as the {\it anomalous dimension\/} of the operator. 

In a CFT the Green functions are covariant under
scale transformations and this provides non-trivial constraints.
As an example, consider the 2-point Green
function $\langle\phi(x)\phi(0)\rangle$. 
This satisfies the more general RG equation \eqref{haq2}
\EQN{
Z(\mu)^{-1}\big\langle\phi(x)\phi(0)\big\rangle_{g_i(\mu),\mu}=Z(\mu')^{-1}
\big\langle\phi(x)\phi(0)\big\rangle_{g_i(\mu'),\mu'}\ .
}
At a fixed point $g_i(\mu)=g_i(\mu')=g_i^*$ and
we have $Z(\mu)=(\mu'/\mu)^{2\gamma_\phi^*}Z(\mu')$, 
where $\gamma_\phi^*=
\gamma_\phi(g_i^*)$. Using dimensional analysis we must have
\EQ{
\big\langle\phi(x)\phi(0)\big\rangle_{g_i^*,\mu}
=\mu^{2d_\phi}{\cal G}(x\mu)\ ,
}
where $d_\phi$ is the classical dimension of the field $\phi$.
Substituting into the RG equation allows us to solve for the unknown
function ${\cal G}$, up to an overall multiplicative constant, yielding
\EQN{
\big\langle\phi(x)\phi(0)\big\rangle_{g_i^*,\mu}
=\frac{c}{\mu^{2\gamma_\phi^*}
x^{2d_\phi+2\gamma_\phi^*}}\propto\frac{1}{
x^{2\Delta_\phi^*}}
}
where $c$ is a constant. This is the typical power-law behaviour
characteristic of correlation functions in a CFT. 
In the problem for this lecture you will see that using the whole of
the conformal group provides even more information.

\BOX{\begin{center}{\bf Relevant, Irrelevant and Marginal}\end{center}

Couplings in the neighbourhood of a fixed point flow as \eqref{RGf}
and are classified in the
following way:\\

(i) If a coupling has
$\Delta_i<d$ the flow diverges away from the fixed point into the IR
as $\mu$ decreases and is therefore known as a {\it relevant\/}
coupling.\\

(ii) If $\Delta_i>d$ the coupling flows into the fixed
point and is known as {\it irrelevant\/}.\\

(iii) The case $\Delta_i=d$ is a marginal
coupling for which one has to go to higher order to find out the
behaviour. If, due to the higher order terms, a coupling
diverges away/converges towards
from the fixed point it is {\it marginally relevant/irrelevant\/}. The
final possibility is that the coupling does not run to all orders. In
this case it is {\it truly marginal\/} coupling and implies that the
original fixed point is actually
part of a whole line of fixed points.
}

When we follow an RG flow backwards towards the UV 
all particle masses decrease relative to cut-off
and the so trajectory of a theory with a mass gap must approach the 
critical surface. In typical cases, with or without a mass-gap,
the trajectory either diverges off to infinity for finite $\mu$ or
approaches a fixed point lying on the critical surface in the limit
$\mu\to\infty$. 
Below we show the RG flows around a fixed point with 2
  irrelevant directions and 1 relevant direction. 
\begin{center}
\includegraphics[width=4in]{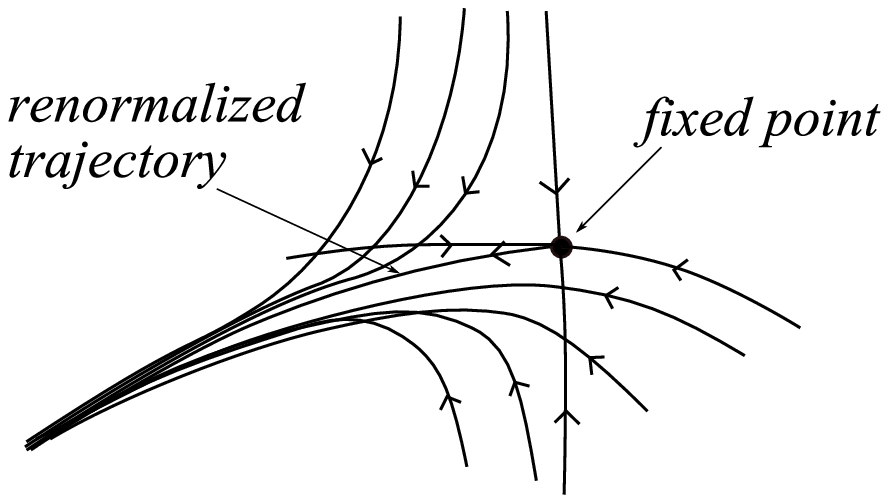}
\end{center}
Notice that the
  flows lying off the critical surface naturally
  focus onto the ``renormalized trajectory'', which is defined as the flow that
  comes out of the fixed point. The focusing effect, along with the fact that
  there are only a finite number of relevant directions, leads to the
  property of {\it Universality\/} 
which is the most important feature of RG flows. Universality also
  arises for flows
  starting on the critical surface since in this case 
they all flow into the fixed point.

\BOX{\begin{center}{\bf Universality}\end{center}

CFTs only have a finite (and usually small) number of relevant
couplings. This means that the domain-of-attraction of a fixed point  (the set
of all points in theory-space) that flow into a fixed point
is infinite dimensional (this
dimension is the number of irrelevant couplings).
This also means that RG flows of a theory lying off the critical
surface strongly focus onto finite
dimensional subspaces parameterized by the relevant couplings of a
fixed point as in the figure above. The
implication of this is that the behaviour of theories in the IR is
determined by only a small number of relevant couplings and not by the
infinite set of couplings $g_i$. This means that IR behaviour of a
given theory with couplings $g_i$ can lie in
a small set of ``universality classes'' which are determined by the
domain of attraction of the set of fixed points.
}

Notice that the RG is directly relevant to the 
problem of taking a continuum limit of a QFT:

\noindent{\sl  The Continuum Limit}

This is the process of taking the cut-off from its original value
$\mu$ to $\infty$ whilst
keeping the keeping the physics at any energy less than the original
$\mu$ fixed. Whether
such a limit exists is a highly non-trivial issue and central to our
story. Notice that taking a continuum limit involves the inverse RG
flow, that is $g_i(\mu)$ with $\mu$ increasing.

The RG Equation
\eqref{haq1} shows how this can be achieved. We can send
$\mu\to\infty$, as long
as the UV limit of $g_i(\mu)$ is suitably well-defined which in practice 
means that $g_i(\infty)$ is a fixed-point of RG (this is what Weinberg
calls ``asymptotic safety). The resulting
$g_i(\mu)$ is known as a ``renormalized trajectory'' since it defines a
theory on all length scales. Clearly a renormalized trajectory
has to have the infinite set of irrelevant couplings at the UV
fixed-point vanishing. Searching for a renormalized trajectory would
seem to involve searching for a needle in an infinite haystack. 
Fortunately, however, universality comes to our rescue: 

\BOX{\begin{center}{\bf Taking a Continuum Limit}\end{center}

We do not need to
actually sit precisely on the renormalized trajectory in order to
define a continuum theory. All that is required is a one-parameter set
of theories defined with cut-off $\mu'$ and with 
couplings $g_i=\tilde g_i(\mu')$ 
(which need {\em not\/} necessarily be an RG flow,
since this would mean sitting on the renormalized trajectory)
for which $\tilde g_i(\infty)$ 
lies in the domain of attraction of the UV CFT, as
illustrated below, where the domain of attraction is the shaded area
\begin{center}
\includegraphics[width=3in]{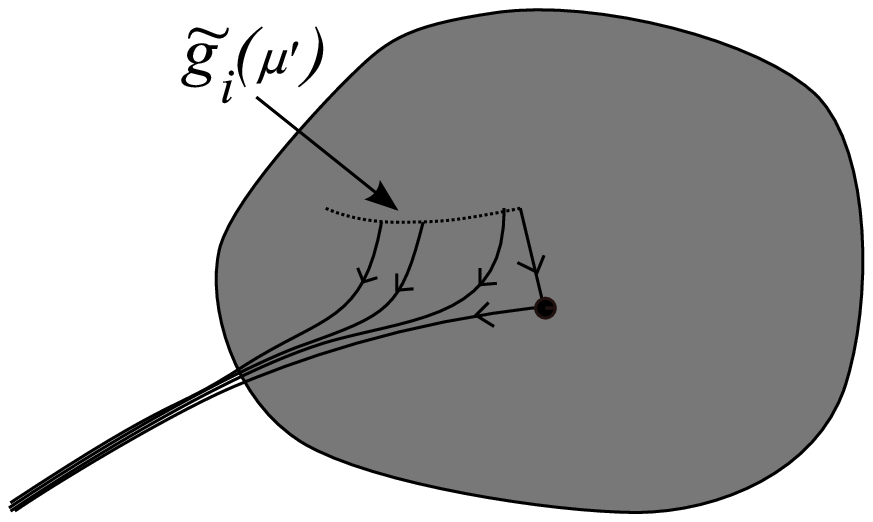}
\end{center}
The limit $\mu'\to\infty$ is defined in such a 
way that the IR physics at the original cut-off scale $\mu$ is fixed. 
In particular, the number of parameters that must be
specified in order to take a continuum limit, {\it i.e.\/}~which fix
the IR physics, equals the number of relevant
couplings of the CFT. However, both the way that relevant couplings are
fixed at the scale $\mu$ and the values of the irrelevant couplings as
$\tilde g_i(\mu')$ as $\mu'\to\infty$ 
can be defined in many different ways. 
So there are many ways to take a continuum limit, or many
``schemes'', which all lead to the same continuum theory. 
In particular, in particle
physics this always allows us to take very simple forms for the action 
with a small number of operators (equal to the number of relevant 
coupling of the UV CFT). However, at the same time it allows us to
describe the same QFT by using, say, a lattice cut-off.
}

\begin{center}{\bf Notes}\end{center}

\begin{quote}
{\small

\begin{note} In the literature,
RG equations like \eqref{haq1} are also often written in infinitesimal
  form as 
\EQN{
\Big[\mu\frac{\partial}{\partial\mu}+\beta_{g_i}\frac{\partial}{\partial
  g_i}\Big]
\FF(g_i;\ell)_\mu=0\ .
}
where $\beta_{g_i}=\mu dg_i/d\mu$.
\label{n11}
\end{note}

\begin{note}The use of the term ``operator'' or ``composite operator''
  derives from a
  canonical quantization approach to QFT in which one builds a Hilbert
  space and on which the fields becomes operators. It is conventional
  to use this language when using the functional integral approach
  when strictly-speaking the quantities are not operators. 
\label{n12}
\end{note}

\begin{note}
You may have noticed that wavefunction renormalization of the fields
in QFT plays a special role relative to the other coupling in the
action. In a sense the definition of RG is ambiguous
since the beta-functions of the couplings depend implicitly on
$Z(\mu)$? Another more concrete way to phrase this 
question is why did we {\it choose\/} the
wavefunction renormalization in order to keep the kinetic term 
intact after RG flow? The intuitive answer is that we want our
theories to describe particles which, at least in the case of a theory
with a mass gap, when well separated 
are approximately free and so have a free field propagator
$\sim i/(p^2-m_\text{ph}^2)$. This is why we re-scaled the field in
the effective action to keep the
kinetic term like that for a free particle. The physical mass of the particle 
$m_\text{ph}$ (which is not the parameter $m$ in the original action) 
is then identified by the position of the pole in the
propagator.

In the statistic physics interpretation of the functional integral,
other choices of wavefunction renormalization are sometimes
appropriate. For instance, in $d\geq4$ one can impose a choice where
the kinetic term is relevant leaving a theory with no propagating
field. In this case, non-zero momentum modes of the field are
suppressed in the IR and only the zero mode, which is constant in
space, survives. This limit is known as {\it mean field theory\/} in
statistical physics.
\label{n112}
\end{note}

\begin{note}Other
  more exotic possibilities, like limit cycles, have been found in
  some very bizarre theories in two spacetime dimensions.
\label{n15}\end{note}

\begin{note}One potential point of confusion regarding fixed points of
  the RG flow is that the fields
    can still have non-trivial wavefunction renormalization factors
    $Z(\phi)$. However, remember that the fields are in some sense
just dummy variables. So a fixed point theory, or CFT, is really scale
    {\it covariant\/} rather than scale {\it invariant}.

The (connected part of the) 
conformal group consists of Poincar\'e transformations along
    with scale transformations, or ``dilatations'' $x\to sx$, and 
special conformal transformations 
\EQN{
x\longrightarrow\frac{x+x^2 b}{1+2b\cdot x+x^2b^2}
}

The infinitesimal transformations for Lorentz, dilatations
and special conformal transformations are
\EQN{
\delta x^\mu=\epsilon^\mu{}_\nu x^\nu\ ,\qquad
\delta x^\mu=sx^\mu\ ,\qquad
\delta x^\mu=x^2b^\mu-2x^\mu(x\cdot b)\ ,
}
where $\epsilon^{\mu\nu}=-\epsilon^{\mu\nu}$.
In any local QFT there exists an energy-momentum tensor $T_{\mu\nu}$
and correlation functions satisfy the Ward identity
\EQN{
\sum_{p=1}^n\big\langle \phi_1(x_1)\cdots \delta\phi_p(x_p)\cdots
\phi_n(x_n)\big\rangle
=-\int d^dx\,\big\langle \phi_1(x_1)\cdots 
\phi_n(x_n)T^\mu{}_\nu(x)\big\rangle\,\partial_\mu\big(\delta x^\nu)\ .
}
Invariance of the QFT under Lorentz transformations requires
$T_{\mu\nu}=T_{\nu\mu}$ while invariance under dilatations
$T^\mu{}_\mu=0$. From this it follows that these two conditions are
sufficient to imply invariance under infinitesimal special conformal
transformations.  
\label{n17}\end{note}

}\end{quote}

\newpage
\section{Scalar Field Theories}

We now illustrate RG theory in the context of the QFT of a single
scalar field. Usually we write down simple actions like
\EQ{
S[\phi]=\int d^dx\,\Big(\frac12\partial_\mu\phi\partial^\mu\phi
-\frac12m^2\phi^2
-\frac1{4!}\lambda\phi^4\Big)\ ,
\label{sac}
}
however, in the spirit of RG we should allow all possible operators
consistent with spacetime symmetries. In the case of a scalar field,
all powers of the field and its derivatives, where the latter are
contracted in a Lorentz invariant way. 
For simplicity we shall restrict 
to operators even in $\phi\to-\phi$.$^{\ref{n20}}$. Simple scaling analysis
shows that a ``composite 
operator'' ${\cal O}$ containing $p$ derivatives
and $2n$ powers of the field, schematically 
$\partial^p\phi^{2n}$, has classical dimension
\EQN{
d_{\cal O}=n(d-2)+p\ .
}
Even at the classical level we see that the number of relevant/marginal
couplings, those with $d_{{\cal O}}\leq d$ is small. The classical
dimensions of various operators are given in the table below.
\begin{center} 
\begin{tabular}{cllll}\toprule
${\cal O}$ & $d>4$ & $d=4$ & $d=3$ & $d=2$ \\ \hline
$\phi^2$ &   rel & rel & rel & rel\\
$\phi^4$ &  irrel & marg & rel & rel\\
$\phi^6$ &  irrel & irrel & marg & rel\\
$\phi^{2n}$ & irrel & irrel & irrel & rel\\
$\partial\phi_\mu\partial^\mu\phi$ & marg & marg & marg & marg\\ 
\toprule
\end{tabular}
\end{center} 

The classical scaling suggests that we only need keep track of the kinetic
term and potential,$^{\ref{n21}}$ 
\EQ{
\LAG=\frac12(\partial_\mu\phi)^2-V(\phi)\ ,
\label{kk3}
}
where we take
\EQN{
V(\phi)=\sum_n\mu^{d-n(d-2)} \frac{g_{2n}}{(2n)!}\phi^{2n}\ ,
}
and where we have used the cut-off to define dimensionless 
couplings $g_{2n}$.

Now we come to crux of the problem, that of finding the RG flow, or
concretely the beta functions, of the
couplings. In order to do this we must apply the RG
equation \eqref{haq3} to the {\it Wilsonian Effective Action\/} 
$S[Z(\mu)^{1/2}\varphi;\mu,g_i(\mu)]$
defined for the theory with cut-off $\mu$ in such a way that the
phenomena on energy scales below the cut-off is fixed as $\mu$ is varied.

Before we can describe how to relate the theories with cut-off $\mu$
and $\mu'$ 
let us first choose a cut-off procedure.
The most basic and conceptually simple way is to introduce a sharp momentum
cut-off on the Fourier modes after Wick rotation to Euclidean
space. In Euclidean space the Lagrangian \eqref{kk3}
has the form
\EQN{
\LAG=\frac12(\partial_\mu\phi)^2+V(\phi)
}
and the functional integral becomes
$\int[d\phi]e^{-S}$.$^{\ref{n212}}$ 
The momentum
cut-off involves Fourier transforming the field
\EQN{
\phi(x)=\int \frac{d^dp}{(2\pi)^d}\tilde\phi(p)e^{ip\cdot x}
}
and then limiting the momentum vector by a sharp cut-off $|p|\leq\mu$. 
The resulting theory is manifestly UV finite since loop integrals can
never diverge. In addition, we have a very concrete way of performing
the RG transformation. Namely, we split the field $\phi$ defined with cut-off
$\mu'$ into 
\EQ{
\phi=\varphi+\hat\phi\ ,
\label{dcp}
}
where $\varphi$ has modes with $|p|\leq\mu$ while $\hat\phi$ are
modes with $\mu\leq|p|\leq \mu'$. 
In order to extract the beta-function it is sufficient 
to consider the infinitesimal transformation
with $\mu'=\mu+\delta\mu$. We can then obtain the RG
flow by considering how the action changes when we 
integrating out $\hat\phi$, so concretely
\EQN{
\exp\big\{-S_\text{eff}[\varphi]\big\}=
\int [d\hat\phi]\,\exp\big\{-S[\varphi+\hat\phi;\mu',
g_{2n}(\mu')]\big\}\ .
}
On the left-hand side we have the {\it Wilsonian Effective Action\/}
which is to be identified with
\EQ{
S_\text{eff}[\varphi]=S[Z(\mu)^{1/2}\varphi;\mu,g_{2n}(\mu)]\ .
\label{ppo}
}
Notice that we have taken $Z(\mu')=1$ since we only need the variation
as $\mu$ changes in order to extract the anomalous dimension $\gamma_\phi$.

Expanding the action on the right-hand side in powers of
$\hat\phi$:$^{\ref{n22}}$
\EQ{
S[\varphi+\hat\phi]=S[\varphi]+\int
d^dx\,\Big(\frac12(\partial_\mu\hat\phi)^2
+
\frac12V''(\varphi)\hat\phi^2+\frac16V'''(\varphi)\hat\phi^3+\cdots\
\Big)\ .
\label{sep}
}

\BOX{\begin{center}{\bf Feynman Diagram Interpretation}\end{center}

Contributions to the effective
action can be interpreted in term of Feynman diagrams with only
$\hat\phi$ on internal lines with a 
propagator $1/(p^2+g_2\mu^2)$, with $p$ integrated over the shell
$\mu\leq |p|\leq \mu'$, and with
only $\varphi$ on external lines (but amputated meaning no propagators
on the external lines). The vertices
are provided by the interaction terms in $V(\varphi+\hat\phi)$. Each loop 
involves an integral over the momentum of $\hat\phi$ which lies in a
shell between radii $\mu$ and $\mu'$ in momentum space:
\EQN{
\int_{\mu\leq|p|\leq\mu'}\frac{d^dp}{(2\pi)^d}\,f(p)\ .
\label{intm}
}
}

However, if we are only interested in an infinitesimal RG
transformation $\mu'=\mu+\delta\mu$ 
then the integrals over internal momenta
\eqref{intm} become much simpler:
\EQN{
\int_{\mu\leq|p|\leq\mu+\delta\mu}\frac{d^dp}{(2\pi)^d} f(p)=\frac{\mu^{d-1}}
{(2\pi)^d}\int d^{d-1}\hat\Omega \,
f(\mu \hat\Omega)\,\delta\mu\ ,
\label{xdd}
}
where $\hat\Omega$ is a unit $d$ vector integrated over a unit $S^{d-1}$. 
In addition, since each loop integral brings a factor of $\delta\mu$,
so to linear
order in $\delta\mu$ only one loop diagrams are needed. This is
equivalent to saying that we only need the term quadratic in $\hat\phi$
in \eqref{sep}. The resulting integral over $\hat\phi$ is Gaussian and
yields$^{\ref{n23}}$
\EQN{
S_\text{eff}[\varphi]=
S[\varphi]+\frac12
\log\det\big(-\square+V''(\varphi)\big)\ .
\label{qwq}
}
In order to extract the RG transformation we identify the
left-hand side with the Wilsonian effective action \eqref{ppo}. In
this case, one finds that no wavefunction 
renormalization is required.$^{\ref{n24}}$ 
In order to extract the effective potential (that is the part of the
Lagrangian not involving derivatives of the field) we can 
temporarily assume that $\varphi$ is constant, in which case
\EQN{
\frac12\log\det\big(-\square+V''(\varphi)\big)
=a\mu^{d-1}\int d^dx\,\log(\mu^2+V''(\varphi))\,\delta\mu\ ,
}
where
$a=\text{Vol}(S^{d-1})/(2(2\pi)^d)=2^{-d}\pi^{-d/2}/\Gamma(d/2)$. 
From this expanding in powers of $\varphi$ it follows that
\BOX{\EQ{
\mu\frac{dg_{2n}}{d\mu}=(n(d-2)-d)g_{2n}-a\mu^{2n}\frac{d^{2n}}{d\varphi^{2n}}
\log(\mu^2+V''(\varphi))\Big|_{\varphi=0}\ .
\label{hya}
}}
The contributions on the right-hand side are identified with one-loop 
diagrams of the form 
\begin{center}
\includegraphics[width=1.5in]{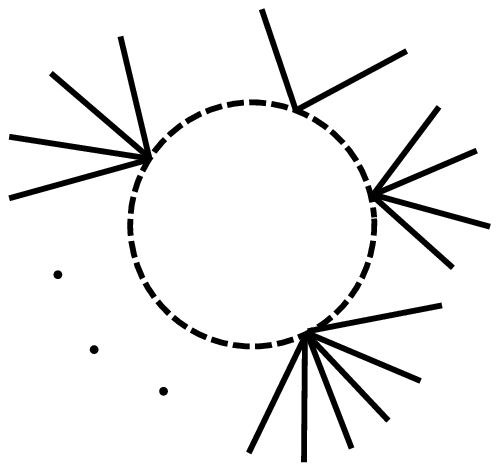}
\end{center}
with $2n$ external legs.

From \eqref{hya}, it follows, for example, that
\SP{
\mu\frac{dg_2}{d\mu}&=-2g_2-\frac{ag_4}{1+g_2}\ ,\\
\mu\frac{dg_4}{d\mu}&=(d-4)g_4+\frac{3ag_4^2}{(1+g_2)^2}-\frac{ag_6}{1+g_2}\ ,\\
\mu\frac{dg_6}{d\mu}&=(2d-6)g_6-\frac{30a g_4^3}{(1+g_2)^3}+\frac{15ag_4g_6}
{(1+g_2)^2}-\frac{ag_8}{1+g_2}\ .\\
\label{sws}
}
Notice that the quantum contributions involve inverse powers of the factor
$1+g_2$ which physically is $m^2/\mu^2+1$. So when $m\gg\mu$,
the quantum terms are suppressed as one would expect on the basis of 
decoupling.

\BOX{\begin{center}{\bf Decoupling}\end{center}

Decoupling expresses the intuition that a particle of mass $m$
  cannot directly affect the physics on energy scales $\ll m$. For
  instance, the potential due to the exchange of massive particle in 4
  dimensions is $\sim e^{-mr}/r$. This is exponentially suppressed
  on distances scales $\gg m^{-1}$.}

The beta functions allow us to map-out RG flow on theory space. 
The first thing to do is to find the RG fixed points corresponding to
the CFTs.
The ``Gaussian'' fixed point is the trivial fixed point where all the
couplings vanish. Linearizing around this point, the
beta-functions are
\EQ{
\mu\frac{dg_{2n}}{d\mu}=\big(n(d-2)-d\big)g_{2n}-ag_{2n+2}\ .
\label{nwn}
}
So the scaling dimensions are the classical dimensions 
$\Delta_{2n}=d_{2n}=n(d-2)$, {\it i.e.}~the anomalous dimensions vanish,
although the couplings that diagonalize the matrix of scaling
dimensions $\sigma_{2n}$ 
are not precisely equal to $g_{2n}$ due to the second term in \eqref{nwn}. In
particular, $\sigma_2=g_2$ is always relevant,
$\sigma_4=g_4+ag_2/(2-d)$ is relevant for $d<4$,
irrelevant for $d>4$ and marginally irrelevant for $d=4$. In this
latter case we need to go beyond the linear approximation. Since
$g_6$ is irrelevant in $d=4$, we shall ignore it, in which case since
$a=1/(16\pi^2)$ we have
\EQN{
\mu\frac{dg_4}{d\mu}=\mu\frac{dg_4}{d\mu}=\frac3{16\pi^2}g_4^2\ ,
}
whose solution is
\EQ{
\frac1{g_4(\mu)}=C-\frac3{16\pi^2}\log\mu\ .
\label{vxx}
} 
This shows that $g_4$ is actually {\it marginally irrelevant\/} 
at the Gaussian fixed point because it gets smaller as $\mu$ decreases.
We usually write the integration constant in terms of a dimensionful
parameter $\Lambda$ as
\EQN{
g_4(\mu)=\frac{16\pi^2}{3\log(\Lambda/\mu)}\ ,
}
with $\mu<\Lambda$. This is our first example of {\it dimensional
  transmutation\/} where the degree-of-freedom 
of a dimensionless coupling $g_4$ has changed
into a dimensionful quantity, namely $\Lambda$.

To find other non-trivial fixed points is difficult 
and the only way we can make
progress is to work perturbatively in the couplings. This turns out to
be consistent only if we accept the perversion of working in arbitrary
non-integer dimension and 
regard $\epsilon=4-d$ as a small parameter. In that case, we find a
new non-trivial fixed point known as the 
Wilson-Fischer fixed point at
\EQ{
g_2^*=-\frac16\epsilon+\cdots\ ,\qquad
g_4^*=\frac1{3a}\epsilon+\cdots\ ,\qquad
g_{2n>4}^*\thicksim \epsilon^n+\cdots
\label{v1}
}
In particular, the Wilson-Fischer fixed point is only physically acceptable if
$\epsilon>0$, or $d<4$, since otherwise the couplings $g_{2n}^*$ are all
negative and the potential of the theory would not be bounded from
below. In the neighbourhood of the fixed point in the $g_2,g_4$
subspace we have to linear order in $\epsilon$
\EQN{
\mu\frac d{d\mu} \MAT{ \delta g_2\\ \delta g_4}
=\MAT{\epsilon/3-2 & -a(1+\epsilon/6)\\ 0 & \epsilon}
 \MAT{\delta g_2\\ \delta g_4}\ ,
}
with 
\EQN{
a=1/(16\pi^2)+\epsilon(1-\gamma_E+\log4\pi)/32\pi^2+{\cal
  O}(\epsilon^2)\ .
}
So the scaling dimensions of the associated operators and the 
associated couplings are
\SP{
&\Delta_2=2-2\epsilon/3\ ,\qquad \sigma_2=\delta g_2\ ,\\
&\Delta_4=4\ ,\qquad \sigma_4=\delta g_4-\frac
a{2+\epsilon/3}\delta g_2 \ .
\label{v2}
}
So at this fixed point only the mass coupling is relevant. 

The flows in the $(g_2,g_4)$ subspace of scalar QFT for small
$\epsilon>0$ are shown below
\begin{center}\includegraphics[width=4in]{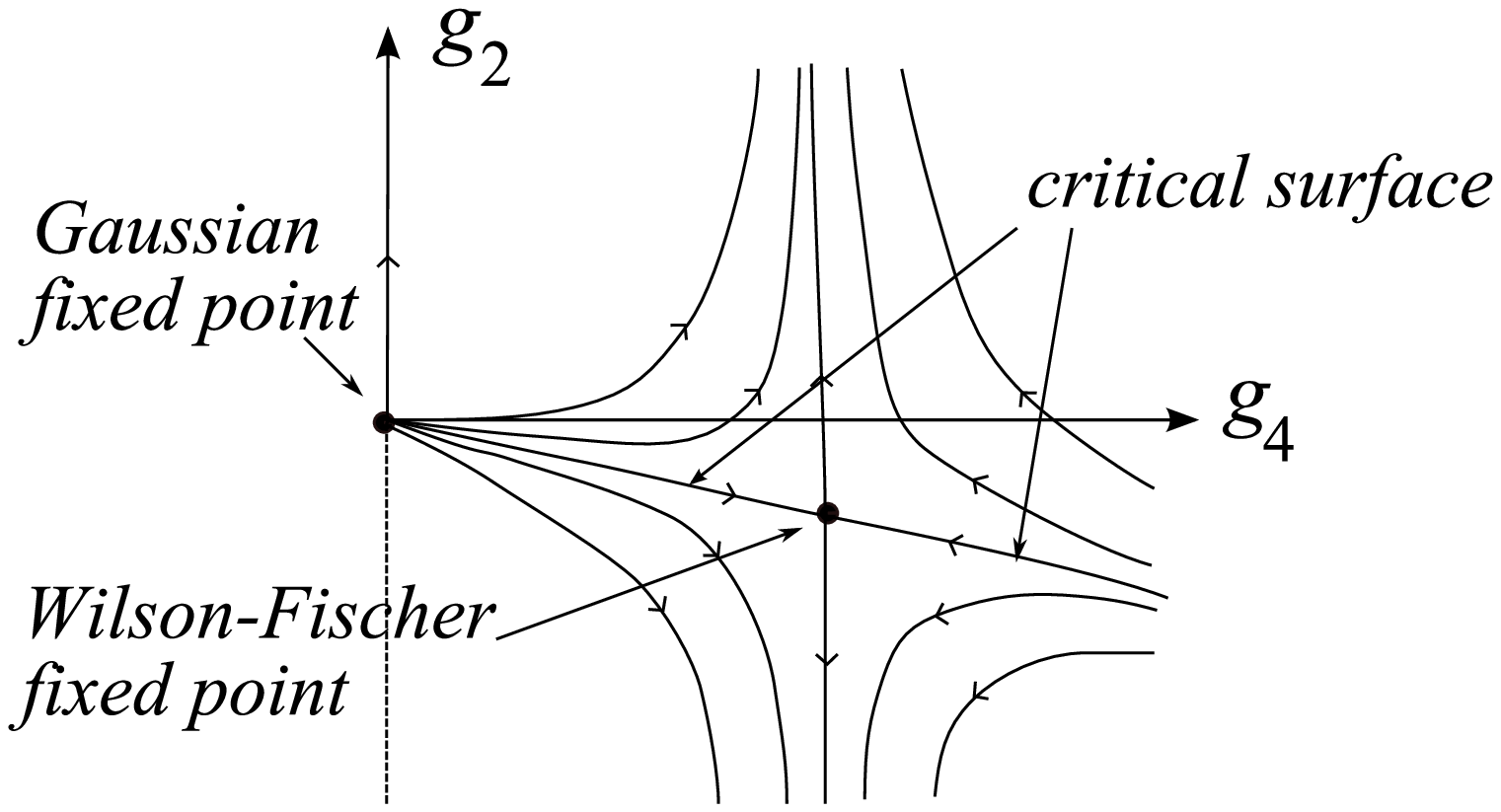}
\end{center}
The Gaussian and Wilson-Fischer fixed points are shown and all
the other couplings are irrelevant and so flow to the $(g_2,g_4)$
subspace. Notice that the
critical surface intersects this subspace in the line that joins the 
two fixed points as shown

Although we have only proved the existence of the Wilson-Fischer fixed
point for small $\epsilon$, it is thought to exist in both $d=3$ and
$d=2$. In the language of statistical physics it lies in
the universality class of the Ising Model.$^{\ref{n241}}$ 
What our simple analysis fails to show is that
in $d=2$ there are actually 
an infinite sequence of additional fixed points.$^{\ref{n25}}$

\BOX{\begin{center}{\bf Vacuum Expectation Values}
\end{center}

In a scalar QFT, the field can develop a non-trivial 
Vacuum Expectation Value (VEV) $\langle\phi\rangle\neq0$. This
possibility is determined by finding the minima of the {\it effective
  potential\/}. This is the potential on the constant (or zero) mode
of the field after all the non-zero modes have been integrated out. In
other words, this is the potential in the Wilsonian effective action
in the limit $\mu\to0$.$^{\ref{n26}}$ A VEV develops when the effective
potential develops minima away from the origin as in
\begin{center}
\includegraphics[width=2in]{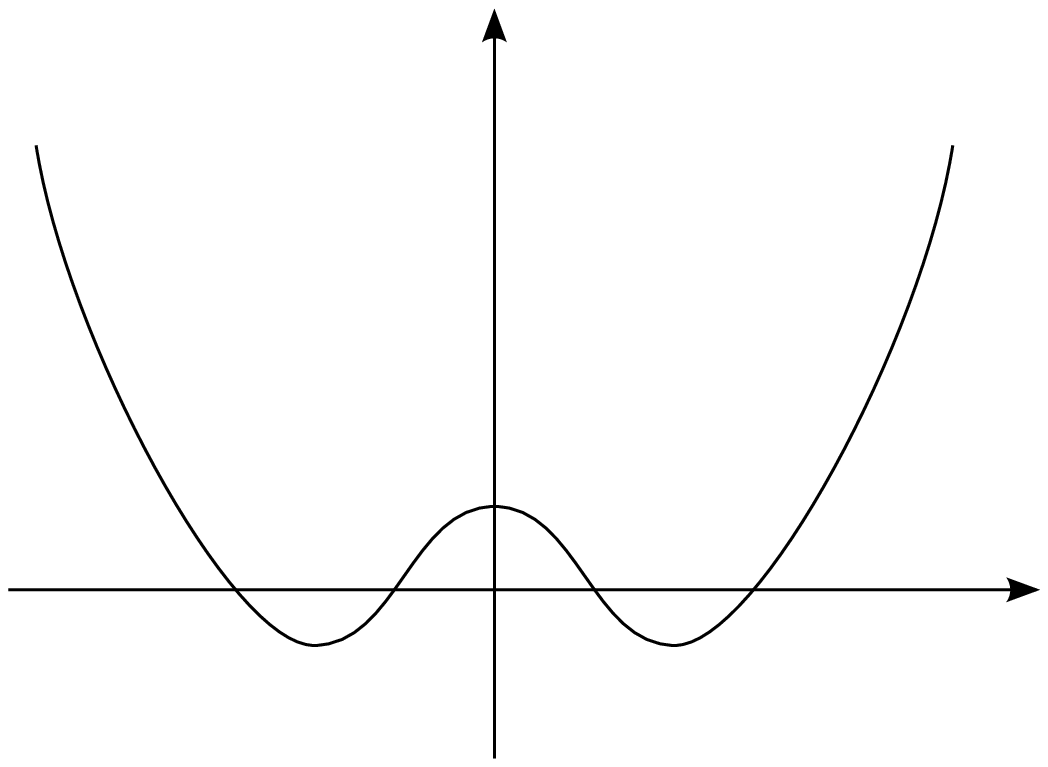}
\end{center}
Notice that since we started with a theory symmetric
under $\phi\to-\phi$, there will necessarily two possible vacuum
states with opposite values of $\langle\phi\rangle$. A QFT must choose
one or the other and so we say that the symmetry $\phi\to-\phi$ is
{\it spontaneously broken\/}.$^{\ref{n27}}$
}

Now that we have a qualitative picture of the RG flows, it is possible
to describe the possible continuum limits of scalar
field theories:

{\bf $d\geq 4$:} In this case, only the
Gaussian fixed point exists and this fixed point only has one relevant
direction, namely the mass coupling $g_2$. Hence there is a single
renormalized trajectory on which
$g_2(\mu)=(\mu'/\mu)^2g_2(\mu')$ while all the other couplings vanish. 
This renormalized trajectory describes the free
massive scalar field. If we sit precisely at the fixed point we have a
free massless scalar field. In particular, according to this analysis
there is no interacting continuum theory in $d=4$.\\ 

{\bf $d<4$:} At least for small enough $\epsilon$ (whatever 
that means) there are two fixed points and a two-dimensional 
space of renormalized
trajectories parameterized by the couplings $g_2$ and $g_4$ 
on which $g_{2n}$, $n>2$ have some values fixed by $g_2$ and $g_4$. 
In particular, if we parameterize our continuum theories by the values
of $g_2$ and $g_4$ then they are limited to the region shown below
\begin{center}
\includegraphics[width=3in]{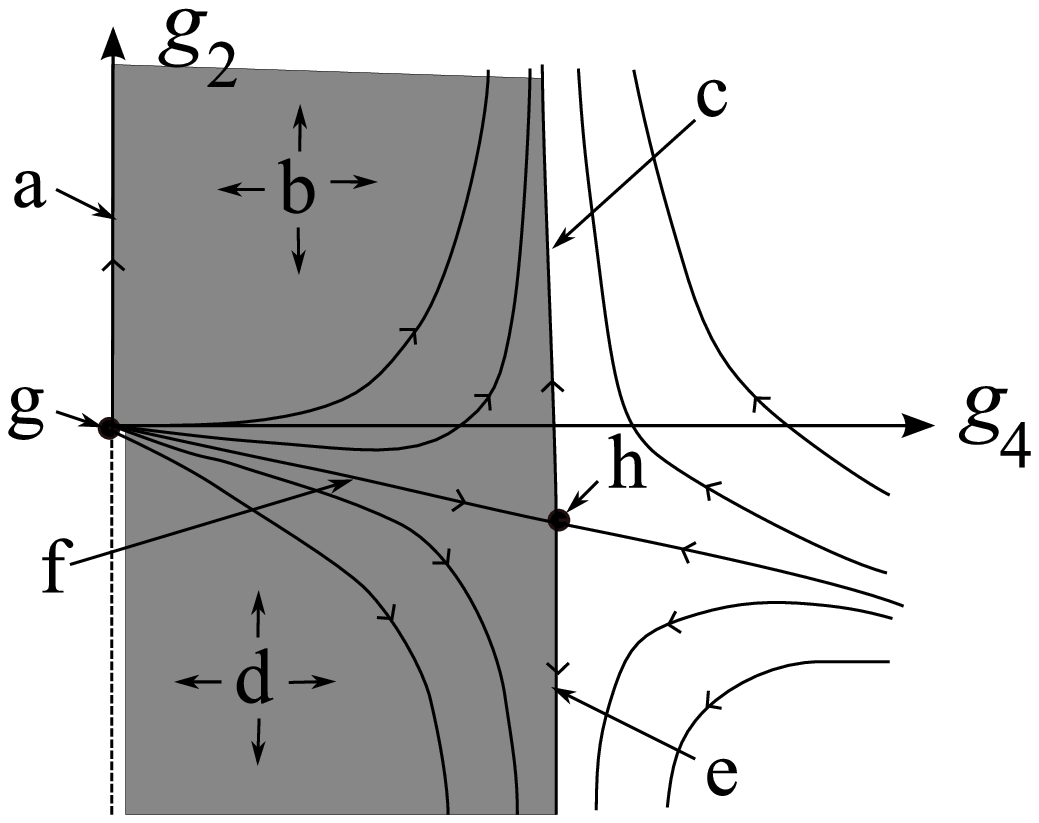}
\end{center}
 In particular, theory (a) is free and massive (and must have
$g_2>0$); (b,d) is interacting and massive and in the UV becomes free
since the trajectories originate from the Gaussian fixed point. In
case $(d)$ $g_2<0$ and the field has a VEV; (c,e)
describe a massive interacting theory that becomes the Ising model
universality class in the UV, case (e) has $g_2<0$ and a VEV; (f)
describes a massless interacting theory that interpolates between a
free theory in the UV and the Ising Model in the IR; (g) is a free
massless theory; and (h) is the Ising model CFT.

\begin{center}{\bf Notes}\end{center}

\begin{quote}
{\small

\begin{note}This is an example of using a symmetry to restrict theory
  space. The important point is that the symmetry is respected by RG
  flow and so is self-consistent.
\label{n20}
\end{note}

\begin{note} Note, however, that, in principle, there is no difficulty
  in keeping track of all the
    couplings including the higher derivative terms.
\label{n21}
\end{note}

\begin{note}We take it as established fact that one can move between
  the Minkowski and Euclidean versions of the theory without
  difficulty. In our conventions, when we Wick rotate $g_{\mu\nu}a^\mu
  b^\nu=a_0b_0-a_1b_1-a_2b_2-a_3b_3\to -a_\mu
  b_\mu=-a_0b_0-a_1b_1-a_2b_2-a_3b_3$.  
In Euclidean space the functional integral $\int
  [d\phi]e^{-S[\phi]}$ can be interpreted as a probably measure (when
  properly normalized) on the the field configuration space. This is
  why Euclidean QFT is intimately related to systems in statistical
  physics.
\label{n212}
\end{note}

\begin{note}In the following we have ignored the term which is linear
  in $\hat\phi$. The reason is that we are after an effective
action for $\varphi$ which is local (that is a spacetime
integral over terms which are powers of the field and its
derivatives). The linear coupling in $\hat\phi$ gives rise to
graphs which are ``1-particle reducible'', the simplest of which is
shown below
\begin{center}
\includegraphics[width=1.5in]{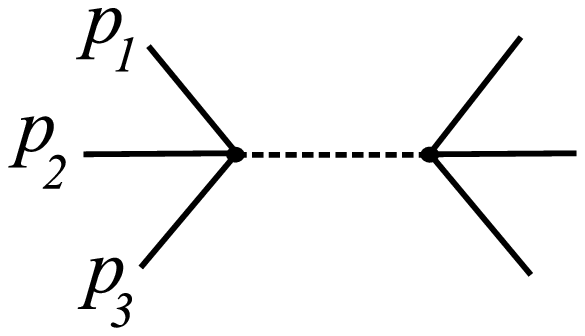}
\end{center}
The contribution from this graph is only non-vanishing if
$\mu\leq|p_1+p_2+p_3|\leq\mu+\delta\mu$. In particular as
long as we work with local actions which are expansions in derivatives, 
{\it i.e.\/}~momenta, we
can ignore such contributions. These non-analytic
contributions are the price we pay for working with a sharp momentum
cut-off. It is important to understand that although the effective
action would have these non-analytic contributions the observables of
the theory, namely the Green functions or the 
S-matrix, would be perfectly well
behaved. For a discussion of this in much greater detail, 
see the excellent lecture notes of Weinberg. 
\label{n22}
\end{note} 

\begin{note} Here we used the identity
\EQN{
\int d^nx\,e^{-x\cdot A\cdot x}=\frac{\pi^{n/2}}{\det\,A}=\pi^{n/2}
e^{-\tfrac12\log\det\,A}
}
for a finite matrix $A$ and extended it to the functional case.
\label{n23}
\end{note}

\begin{note}The terms quadratic in $\phi$ come from the diagram
below
\begin{center}
\includegraphics[width=0.8in]{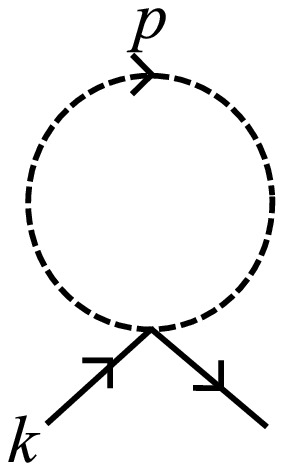}
\end{center}
which gives a contribution
\EQN{
\thicksim \int \frac{d^dk}{(2\pi)^d} \tilde \phi(k)\tilde \phi(-k)
\,\int \frac{d^dp}{(2\pi)^d}\frac1{p^2+g_2\mu^2}
=\text{const.}\times\int d^dx\,\phi(x)^2\ .
}
Notice that the resulting expression does not involve 
derivatives of the field because, 
 due to momentum conservation, no external momentum can flow 
in the loop. Hence, there is no wavefunction renormalization.
\label{n24}
\end{note} 

\begin{note}The Ising Model is a statistical model defined on a square
  lattice with spins $\sigma_i\in\{+1,-1\}$ at each site and with an
  energy (which we identify with the Euclidean action)
\EQN{
\mathscr{E}=-\frac1T\sum_{(i,j)}\sigma_i\sigma_j\ .
}
The sum is over all nearest-neighbour pairs $(i,j)$ and $T$ is the
  temperature. Notice that at low temperatures, the action/energy
  favours alignment of all the spins, while at high temperatures
  thermal fluctuations  are large and the long-range order is
  destroyed. This can be viewed as a competition between energy and
  entropy. There is a 2nd order phase transition at a critical
  temperature $T=T_c$ at which there are long-distance power-law
  correlations. This critical point is in the same universality class
  as the Wilson-Fischer fixed point and the water-steam critical point.
\label{n241}
\end{note}

\begin{note}In $d=2$ there are powerful methods for analyzing CFTs
  because in $d=2$ the conformal group is infinite dimensional since
  it consists of any holomorphic transformation $t\pm x\to f(t\pm x)$.
\label{n25}
\end{note}

\begin{note}It can be shown that the effective potential defined in
  terms of the Wilsonian effective action is equal to the 
effective potential extracted from the more familiar 
1-Particle Irreducible (1-PI) effective
  action defined in perturbation theory---at least for a QFT with a mass gap. 
In the massless case the latter quantity
  is ill-defined due to IR divergences.
\label{n26}
\end{note}

\begin{note}The reason why spontaneous symmetry breaking occurs is
  that the zero mode of a scalar field is not part of the variables
  that are integrated over in the measure $\int[d\phi]$, rather it acts as
  a boundary condition on the scalar field at spatial infinity. 
However, this is only true
  in spacetime dimensions $d>2$: in $d=2$ one must integrate over the
  zero mode and so spontaneous symmetry breaking of {\it continuous\/}
  symmetries cannot occur. Of course $\phi\to-\phi$ is a discrete
  symmetry that can still be spontaneously broken even in $d=2$.
\label{n27}
\end{note}

}\end{quote}

\newpage
\section{RG and Perturbation Theory}

In this lecture, we consider how to implement RG ideas in the context
of perturbation theory and in a way which is
easier to generalize to other theories.

It is the principle of 
universality that allows us to formulate our theories in terms
of simple actions. All we need do is include the relevant
couplings: all the irrelevant couplings can be taken to vanish. 
So---at least for $d>2$---it is sufficient to write the simple action
in \eqref{sac}.$^{\ref{n31}}$ In order to take a continuum limit, we need
to let the ``bare couplings'' $g_i(\mu')$ to depend on the cut-off $\mu'$
in such a way that the physics at a physically relevant scale 
$\mu<\mu'$ remains fixed. This is guaranteed if we use the 
RG equation \eqref{haq1} and 
take $\mu'\to\infty$ with the couplings $g_i(\mu')$
following the RG flow into the UV. However, this would mean
keeping all the couplings to irrelevant operators. As we described in
Lecture 1, there is no need to do this. It is sufficient to keep only
the relevant couplings $g_2(\mu')$ and $g_4(\mu')$ (as well as the kinetic
term) and choose all the irrelevant couplings to vanish. So in this
sense, the continuum limit flow is not strictly speaking an RG flow;
rather it is the RG flow restricted to the space of relevant couplings. 

\BOX{\begin{center}{\bf Perturbation Theory: The Facts of
      Life}\end{center}

Now we come to a very important issue. In real life 
we have to rely almost always 
on perturbation theory in the couplings. But
according to the RG the couplings flow and this begs the question as 
to which coupling should use to perform
perturbation theory? For example in $d=4$, $g_4$ runs according to
\eqref{vxx}. So the coupling at the scale $\mu'$ 
is an infinite perturbative
series of the coupling at $\mu$:
\EQ{
g_4(\mu')=\frac{g_4(\mu)}{1+\tfrac3{16\pi^2}g_4(\mu)\log\mu'/\mu}
=g_4(\mu)-\frac3{16\pi^2}g_4(\mu)^2\log\mu'/\mu+\cdots\ .
\label{qll}
}
If we could completely sum all of the perturbative expansion in
$g_4(\mu)$, then the resulting physical observables would---by 
definition---be independent of $\mu$. However, 
since perturbation theory involves a truncation it is clear that
perturbation theory in $g_4(\mu)$ will be different from perturbation
theory in $g_4(\mu')$. Do we have a choice of $\mu$ and can we choose
$\mu$ so that $g_4(\mu)$ is small and perturbation theory is reliable? 
The answer is practically 
no: the appropriate choice of $\mu$ is dictated by the
characteristic physical scale involved and we just have to hope that
at this scale $g_4(\mu)$ is small. If we try to do perturbation at a
non-physical scale then any improvement we might make in having a
smaller $g_4(\mu)$ is compensated by larger coefficients multiplying
the powers of the coupling.
We will see an example of how the physical scale dictates the choice
of $\mu$ in a procedure
known as {\it renormalization group improvement\/}.
}

So we should perform 
the perturbation expansion in the coupling at the physically relevant 
 energy scale $\mu$, {\it i.e.\/}~in the ``renormalized coupling'' 
$\lambda\equiv\lambda(\mu)$ (we choose to use $\lambda$ and $m$ rather
than $g_4$ and $g_2$ in the following description)
rather than
the ``bare coupling'' $\lambda_b\equiv\lambda(\mu')$ since $\lambda(\mu)<
\lambda(\mu')$. To this end, we split up the bare Lagrangian 
into the ``renormalized'' part and the ``counter-term'' part:
\EQN{
\LAG=\frac12(\partial_\mu\phi_b)^2+\frac12m_b^2\phi_b^2
+\frac1{4!}\lambda_b\phi_b^4
=\LAG_r+\LAG_{ct}
}
where $\LAG_r$ has the same form as the bare Lagrangian, 
but with bare quantities replaced by renormalized ones (which we denote
$\phi$, $m$ and $\lambda$):
\EQN{
\LAG_r=\frac12(\partial_\mu\phi)^2+\frac12m^2\phi^2
+\frac1{4!}\lambda\phi^4
}
and $\LAG_{ct}$, the counter-term Lagrangian,
\EQN{
\LAG_{ct}=\frac12\delta Z(\partial_\mu\phi)^2+\frac12\delta m^2\phi^2
+\frac1{4!}\delta\lambda\phi^4\ .
}
such that the bare quantities are
\EQ{
m^2_b=m^2+\delta m^2\ ,~~~~
\lambda_b=\lambda+\delta\lambda\ ,~~~~\phi_b=\sqrt{1+\delta Z}\phi\ .
}
It is important to emphasize that although we have denoted the
counter-terms as $\delta\#$ they certainly are {\it not\/} infinitesimals.

Perturbation theory can be be performed in the renormalized coupling 
$\lambda$ and the counter-terms found order-by-order in $\lambda$
without prior knowledge of the RG flow. In this point-of-view, the
counter-terms are chosen to cancel the divergences that occur in the
limit $\mu'\to\infty$.$^{\ref{n33}}$ For instance, in $d<4$, the
only Feynman diagrams which are superficially divergent$^{\ref{n34}}$ have 2
external legs and are shown below:
\begin{center}
\includegraphics[width=3in]{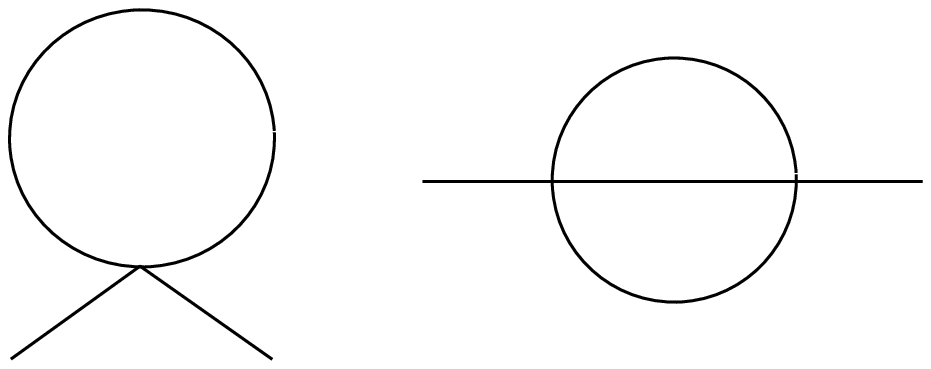}
\end{center}
While in $d=4$ (and also for small $\epsilon=4-d$) all diagrams with 2
and 4 external lines are superficially divergent.

While we could continue with the sharp momentum cut-off, it is time to
acknowledge its frailties. It has been a useful device 
to introduce the concept of 
RG flow, however, when we start to investigate gauge theories we
discover that it is not obvious how to make 
the na\"\i ve sharp momentum cut-off
consistent with gauge invariance. Fortunately, our all-dimension 
treatment of the scalar
field naturally leads to the most important regularization scheme---at 
least within perturbation theory---known as {\it dimensional
  regularization\/}. The idea is to use the fact that we have defined
the theory in $d=4-\epsilon$ dimensions and treat $\epsilon$ as a
variable in its own right. Then, after performing the loop integrals, we
will have analytic functions of $\epsilon$. The divergences in the
physical dimensions $d=2,3,4\,
\ldots$ show up as poles as $\epsilon\to2,1,0$.
For example, in four dimensions we make the replacement
\EQN{
\int\frac{d^4p}{(2\pi)^4}\longrightarrow \mu^{4-d}\int\frac{
d^dp}{(2\pi)^d}\ ,
}
where $\mu$ is a parameter, with unit mass dimension, which
is introduced to make sure that the momentum integrals have the
correct dimension. We will see that in a subtraction scheme known as
{\it minimal subtraction\/} it plays a r\^ole analogous to the
Wilsonian cut-off scale $\mu$ which is the scale at which the 
renormalized coupling are defined.

As an example consider the quadratically divergent one-loop graph in $d=4$ 
\begin{center}
\includegraphics[width=1in]{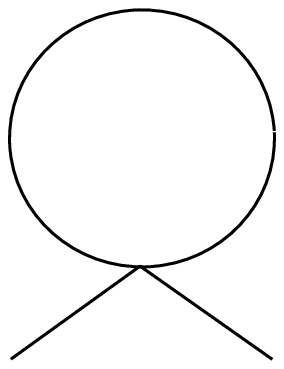}
\end{center}
which involves the loop integral
\SPN{
\lambda\mu^{4-d}\int\frac{d^dp}{(2\pi)^d}\cdot
\frac1{p^2+m^2}&=\lambda\mu^{4-d}\frac{\text{Vol}S^{d-1}}{
(2\pi)^d}\int_0^\infty\frac{p^{d-1}dp}{p^2+m^2}\\
&=\lambda 2^{-d}\pi^{-d/2}\mu^{4-d}m^{d-2}\Gamma(1-d/2)\ .
}
In the limit $\epsilon=4-d\to0$ this equals
\EQ{
-\frac{\lambda m^2}{8\pi^2\epsilon}+\frac{\lambda m^2}{16\pi^2}\Big[-1
+\gamma_E-\log\frac{4\pi \mu^2}{m^2}\Big]+{\cal O}(\epsilon)\ .
}
The divergences can be removed by a mass
counter-term of the form
\EQ{
\delta m^2=\frac{\lambda m^2}{16\pi^2\epsilon}+\frac{\lambda
  m^2}{32\pi^2}\big(-\gamma_E+\log4\pi\big)\ .
\label{nen}
}
As we have previously stated there is considerable freedom in removing
the divergences. The choice \eqref{nen} is known as {\it modified minimal
subtraction\/}, denoted \MS. Here ``minimal'' refers to the singular
part in \eqref{nen} 
and ``modified'' refers to the second and third terms which are
optional extras but part of the perturbative industry standard. 

The other divergent diagram at one-loop is 
\begin{center}
\includegraphics[width=1.8in]{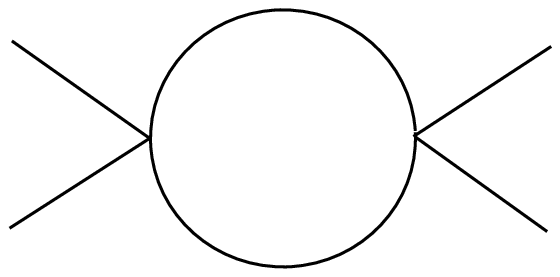}
\end{center}
In this case the loop
integral is logarithmically divergent in $d=4$:
\EQN{
\lambda^2\mu^{4-d}\int\frac{d^dp}{(2\pi)^d}\cdot
\frac1{p^2+m^2}\cdot\frac1{(p+k)^2+m^2}\ ,
}
where $k=k_1+k_2$. Expanding in terms of the external
momentum $k$ it is only the term $k=0$ which is divergent since every
power of $k$ effectively gives one less power of $p$ for large
$p$. The divergent piece is then
\SPN{
\lambda^2\mu^{4-d}\int\frac{d^dp}{(2\pi)^d}\cdot\frac1{(p^2+m^2)^2}
&=\lambda^2\mu^{4-d}\,\frac{\text{Vol}\,S^{d-1}}{(2\pi)^{d}}
\int_0^\infty\frac{p^{d-1}dp}{(p^2+m^2)^2}\\
&=\lambda^22^{-d}\pi^{-d/2}\mu^{4-d}m^{d-4}\Gamma(2-d/2)\\
&=
\frac{\lambda^2}{8\pi^2\epsilon}+\frac{\lambda^2}{16\pi^2}
\Big[-\gamma_E+\log\frac{4\pi\mu^2}{m^2}\Big]+{\cal O}(\epsilon)\ .
}
In minimal subtraction the divergence can be cancelled by the counter-term
\EQ{
\delta\lambda=\frac{3\lambda^2}{16\pi^2\epsilon}+\frac{3\lambda^2
 }{32\pi^2}\big(-\gamma_E+\log4\pi\big)\ .
\label{nen2}
}
Later we shall see that regularization schemes like minimal
subtraction are very different in character 
from alternative and more physically
motivated schemes.$^{\ref{mc1}}$

As mentioned above, the renormalized couplings will depend on the energy scale
$\mu$ in a way that is analogous to $\mu$ of the sharp momentum
cut-off scheme. 
In fact in the we can recover the spirit of Wilson's
RG by using what is known as the {\it background field
  method\/}. The idea is to expand the field $\phi$ in 
the renormalized action in 
terms of a slowly varying background field and a more rapidly fluctuating part:
\EQ{
\phi=\varphi+\hat\phi\ .
}
This is analogous to the decomposition \eqref{dcp} in the sharp
momentum cut-off scheme. We then treat $\hat\phi$ as the 
field to integrate over in the functional integral
whilst treating $\varphi$ as a fixed background field.  

Since $\varphi$ is slowly varying we can expand in powers of derivatives of
$\varphi$ and as in Lecture 2, 
and in order to calculate the effective potential we can take
$\varphi$ to be constant. The effective potential in $d=4$ to one loop is then
\SPN{
V_\text{eff}(\varphi)&=V(\varphi)+\frac12\mu^{4-d}
\int \frac{d^dp}{(2\pi)^d}\,\log\big(p^2+V''(\varphi)\big)\\
&=V(\varphi)-\frac{\big(V''(\varphi)\big)^2}{32\pi^2\epsilon}-\frac{
\big(V''(\varphi)\big)^2}{64\pi^2}\Big[\frac32-\gamma_E+\log
\frac{4\pi\mu^2}{V''(\varphi)}\Big]\ .
}
Subtracting the divergence with a counter-term 
in the $\overline{\text{MS}}$ scheme gives
\EQ{
V_\text{eff}(\varphi)=V(\varphi)
-\frac{
\big(V''(\varphi)\big)^2}{64\pi^2}\Big[\frac32+\log
\frac{\mu^2}{V''(\varphi)}\Big]\ .
\label{vef}
}
With a potential $V(\phi)=\tfrac12m^2\phi^2+\tfrac1{4!}\lambda\phi^4$,
notice that the counter-term is proportional to
\EQ{
\big(V''(\varphi)\big)^2=\big(m^2+\tfrac12\lambda\varphi^2\big)^2
}
which has the same form as the original
potential. In other words, we only need to add counter-terms for $m^2$
and $\lambda$. With a little more work one finds that the 
counter-terms are precisely \eqref{nen} and \eqref{nen2}.

The RG flow of the dimensionless 
couplings $g_2=m^2/\mu^2$ and $g_4=\lambda$ can be deduced, as before, by 
requiring that the effective potential satisfies an RG equation. For
instance, in the form \eqref{haq3},
\EQN{
V_\text{eff}(Z(\mu)^{1/2}
\varphi;\mu,g_i(\mu))=V_\text{eff}(Z(\mu')^{1/2}\varphi;\mu',
g_i(\mu'))\ ,
}
where now $\mu$ is no longer the Wilsonian cut-off but is the mass scale
introduced in dimensional regularization to ``fix up'' the
dimensions. 

To one-loop order there is no wavefunction renormalization
and using the above, we find
\SP{
\mu\frac{dg_2}{d\mu}&=-2g_2+ag_2g_4\ ,\\
\mu\frac{dg_4}{d\mu}&=3ag_4^2\ .
\label{gwg}
}
where $a=1/16\pi^2$. Compare \eqref{gwg} with the momentum
cut-off scheme \eqref{sws} (with $d=4$ and with the 
 irrelevant couplings set to 0). The first difference is that with the
latter scheme the beta-function is exact at the one-loop level,
whereas \eqref{gwg} receives contributions to all loop
orders. Secondly, the momentum cut-off scheme displays manifest decoupling
when $m\gg\mu$, whereas, on the contrary, dimensional regularization
with $\overline{\text{MS}}$ subtraction does not. In the next lecture
we will explore this in more detail and explain how decoupling must be
implemented by hand in the  $\overline{\text{MS}}$ scheme.
What these two
schemes illustrate is that the actual RG flows depend in detail on 
the chosen scheme. However, it is a central feature of the theory that the
``topological'' properties of the flows, meaning existence of fixed
points, or whether a coupling is relevant or irrelevant, is scheme-independent.
For example, if we work in dimensions $d<4$ rather than $d=4$, then one
can demonstrate the existence of the Wilson-Fischer fixed point.

Our analysis of scalar field theories leads to the following important
conclusion for spacetime dimension $d=4$. Since we have only proved the
existence of the Gaussian fixed point, the only continuum
theory is a free massive or massless 
theory. This lack of an interacting continuum
scalar theory in 4 dimensions is known as {\it
  triviality\/}. But, you will say, haven't we been able to find a
renormalizable theory in 4 dimensions order-by-order in the
perturbative expansion: isn't this an interacting theory?
However, something does go terrible wrong with perturbation theory and
this is caused by the wish to keep the renormalized coupling finite
whilst sending the cut-off $\mu'\to\infty$. We can see what
goes wrong by looking at the
flow of the coupling \eqref{qll}: we see that
as $\mu'$ increases there is singularity at
\EQN{
\mu'=\mu e^{16\pi^2/3g_4(\mu)}\ ,
}
where the bare coupling diverges. This is known as 
a {\it Landau pole} (this terminology is properly associated with QED which,
as we shall see, suffers the same fate). Of course this is exactly what
we expected: the coupling $g_4(\mu)$ is irrelevant at the Gaussian
fixed point and so RG flow into the UV diverges away from the fixed
point. Now we see within the one-loop approximation that the flow
actually diverges to infinity at a finite value. So the conclusion is, 
in the absence of a non-trivial fixed point,
scalar QFT is not really a renormalizable theory in $d=4$. 
This conclusion applies more or less to the scalar Higgs sector
of the standard model, and so in this sense the standard model is not
truly a renormalizable theory and so, in a sense, predicts its own demise.

\BOX{\begin{center}{\bf Summary}\end{center}
 
We now sketch how to use perturbation theory (when valid) and
the background field method
to calculate the RG flows of the revelant couplings.\\

(1) Write down a Lagrangian $\LAG_r(\phi)$ 
with all the ``relevant'' couplings $g^r_i$. \\

(2) Split $\phi=\varphi+\hat\phi$ and 
calculate loop diagrams with external $\varphi$ and internal
    $\hat\phi$ (more precisely only include diagrams which are
    {\it one particle irreducible\/} 1PI)
and add counter-terms along the way to cancel divergences. It is
important that the counter-terms have the same form as $\LAG_r(\varphi)$
(if not then you didn't identify the complete set of relevant couplings).
\\

(3) The analogue of the Wilsonian effective action in the special
subspace of theories parameterized by the relevant coupling is then 
\EQ{
S[\varphi;\mu,g^r_i]=\int d^dx\,\Big(\LAG_r(\varphi)+\LAG_{ct}(\varphi)+
\LAG_\text{1PI}(\varphi)\Big)\ .
}
Note that this action will include all the irrelevant
couplings as well, but these will not be independent rather they 
will depend on the relevant couplings.
\\

(4) Extract the RG flows of the revelant couplings in $\LAG_r(\varphi)$  by
    imposing the RG equation \eqref{haq3}.
}

\begin{center}{\bf Notes}\end{center}

\begin{quote}
{\small

\begin{note}In $d=2$ the situation is more complicated and we have to
allow for an arbitrary potential since all operators $\phi^n$ are
potentially relevant.
\label{n31}\end{note}

\begin{note}There is considerable freedom in choosing the way that the 
divergent terms are cancelled. These choices should be thought of as
part of the regularization {\it scheme\/}.
\label{n33}\end{note}

\begin{note} 
The question of whether a given diagram is divergent can
  be partially addressed by calculating the {\it superficial degree of
  divergence\/} ${\cal D}$. This is the power of the 
overall momentum dependence of
  the diagram: each propagator contributes $\sim p^{-2}$ and each loop integral
  $\sim p^d$. One can show for a $\phi^4$ theory that for a diagram
  with $L$ loops and $E$ external lines
\EQN{
{\cal D}=(d-4)L+4-E\ .
}
Notice that when ${\cal D}\geq0$ the diagram is divergent, however,
  the converse doesn't imply convergence since a sub-diagram may be
  superficially divergent. When a theory only has 
superficial divergences in a finite number of
  diagrams it is called {\it super-renormalizable\/}.
For small $\epsilon=4-d$ we expand first in $\epsilon$ and
  later take the continuum limit $\mu\to\infty$. In this case, the
  superficially divergent diagrams include all those 2 or 4 external
  legs. 
\label{n34}\end{note}

\begin{note}\MS\ subtraction is not the
  only way to remove the divergences in dimensional regularization. A
  more physically motivated scheme 
involves fixing the physical mass
  and physical coupling defined in terms of the propagator and 
Green functions at some characteristic Euclidean momentum scale
  $\tilde\mu$:
\SPN{
&\langle\phi(p_1)\phi(p_2)\rangle=\frac1{p_1^2+m^2}\delta^{(4)}(p_1+p_2)
\ ,\\
&\langle\phi(p_1)\phi(p_2)\phi(p_3)\phi(p_4)\rangle=
\lambda\delta^{(4)}(p_1+p_2+p_3+p_4)\ ,
}
when $p_i^2=\tilde\mu^2$. If we Wick rotate back to Minkowski space these
  conditions occur at space-like momenta $p_i^2=-\tilde\mu^2$.

In this kind of momentum-dependent scheme, the quantity $\tilde\mu$ 
plays the r\^ole of the RG scale. The
counter-terms $\delta g_i$ depend explicitly on $\tilde\mu$ and the beta
functions are equal to
\EQ{
\tilde\mu\frac{d\delta g_i}{d\tilde\mu}\ ,
\label{pxx}
}
where as usual the $g_i$ are made dimensionless by using $\tilde\mu$. It is
important to realize, as we shall see in the next lecture, that the
beta-functions in this scheme are very different from the ones in the
\MS\ scheme.
\label{mc1}\end{note}

}\end{quote}

\newpage

\section{Gauge Theories and Running Couplings}

In this lecture, we turn our attention to the RG properties of 
gauge theories. We begin with the simplest gauge theory; namely,
QED. In order that QED has interesting RG flow we require it to be
interacting and to achieve this end we couple it to a charged 
scalar. The scalar field must necessarily be complex and for
simplicity we
take a (Minkowski) Lagrangian without any self-interactions for $\phi$:
\EQ{
\LAG=
-\frac14F_{\mu\nu}F^{\mu\nu}+|D_\mu\phi|^2-m^2|\phi|^2\ ,
}
where $D_\mu\phi=\partial_\mu\phi+ieA_\mu\phi$ and
$F_{\mu\nu}=\partial_\mu A_\nu-\partial_\nu A_\mu$. The theory is
invariant under gauge transformations
\EQ{
\phi(x)\to e^{ie\alpha(x)}\phi(x)\ ,\qquad
A_\mu(x)\to A_\mu(x)-\partial_\mu\alpha\ .
}

We want to focus
on the RG flow of the charge $e$. In order to do this it is actually
more convenient to re-scale $A_\mu\to A_\mu/e$ so that the coupling
now appears in front of the photon's kinetic term
\EQN{
\LAG=-\frac1{4e^2}F_{\mu\nu}F^{\mu\nu}
+|(\partial_\mu+iA_\mu)\phi|^2-m^2|\phi|^2\ .
}
The reason why this is a
convenient choice in the context of RG is that under the flow the structure
of the covariant derivative must remain intact otherwise gauge
invariance would not be respected.
This means that $A_\mu$ with the Lagrangian as above must undergo no  
wavefunction renormalization, rather we interpret any renormalization
of the gauge kinetic term as a renormalization of the electric charge 
$e$.$^{\ref{vp1}}$ 

In order to find the RG flow of the coupling we use the background
field method and consider $A_\mu$ as a background field and 
treat $\phi$ as the fluctuating field that is to be integrating out in the
functional integral. The interaction terms are
\EQ{
\LAG_\text{int}=iA_\mu\big(\phi\partial^\mu\phi^*-\phi^*\partial^\mu\phi\big)
+A_\mu A^\mu|\phi|^2\ .
}
At the one-loop level there are 2 relevant graphs shown below
\begin{center}
\includegraphics[height=0.7in]{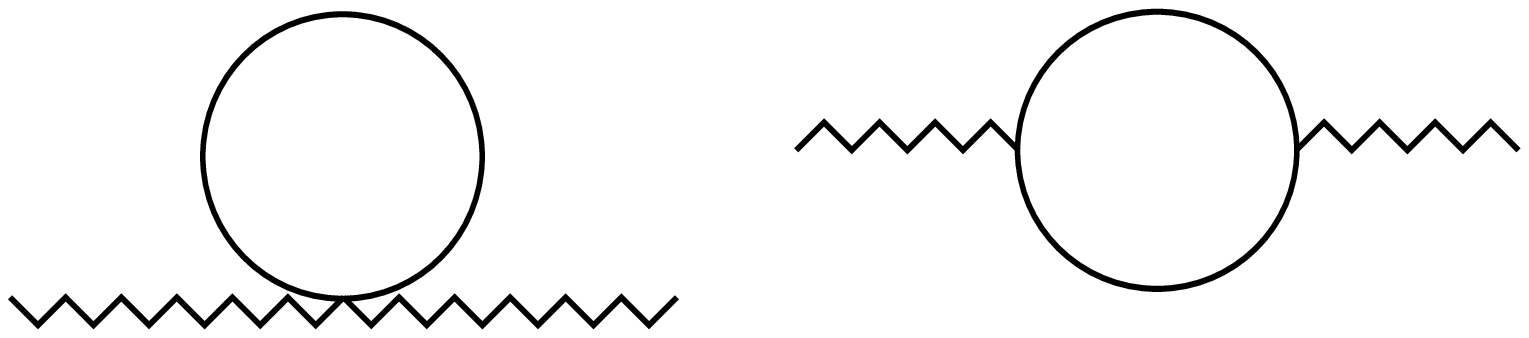}
\end{center}
The first one cannot depend on the external momentum $k$
and its r\^ole is to cancel the $k$-independent contribution of the
first. In dimensional regularization, and temporarily Wick rotating to
Euclidean space, the second diagram yields
\EQ{
\mu^{4-d}\int\frac{d^dp}{(2\pi)^d}\cdot\frac{(2p+k)_\mu(2p+k)_\nu}
{\big(p^2+m^2\big)\big((p+k)^2+m^2\big)}\ .
\label{q10}
}
The standard trick for dealing with a product of propagators is to
introduce Feynman parameters, which in this case involves the identity 
\EQN{
\frac1{p^2+m^2}\cdot\frac1{(p+k)^2+m^2}=\int_0^1dx\,\frac1{\big((p+xk)^2+x(1-x)
k^2+m^2\big)^2}
}
and then shifting $p\to p-xk$, 
in which case \eqref{q10} becomes
\SPN{
&\mu^{4-d}\int_0^1dx\,
\int\frac{d^dp}{(2\pi)^d}\cdot\frac{(2p+(1-2x)k)_\mu(2p+(1-2x)k)_\nu}
{\big(p^2+x(1-x)
k^2+m^2\big)^2}\\
&=\mu^{4-d}\frac{\Gamma(1-d/2)}{(4\pi)^{d/2}}\int_0^1dx\,\Delta^{d/2-2}\Big[
(1-d/2)(1-2x)^2k_\mu k_\nu+2\Delta\delta_{\mu\nu}\Big]
}
where $\Delta=x(1-x)k^2+m^2$. The second term in the bracket can be
re-written using the identity
\EQN{
2\int_0^1dx\,\Delta^{d/2-1}=k^2(d/2-1)\int_0^1dx\,(1-2x)^2
\Delta^{d/2-2}+2m^{d-2}\ .
}
The last term here cancels the contribution 
\EQN{
\mu^{4-d}\int\frac{d^dp}{(2\pi)^d}\,\frac{\delta_{\mu\nu}}{p^2+m^2}
=\mu^{4-d}\frac{\Gamma(1-d/2)}{(4\pi)^{d/2}}m^{d-2}
}
from the first diagram above. This leaves
\EQ{
\frac{\mu^{4-d}}{(4\pi)^{d/2}}\big(k_\mu
  k_\nu-k^2\delta_{\mu\nu}\big)
\Gamma(2-d/2)\int_0^1dx\,(1-2x)^2\big(x(1-x)k^2+m^2\big)^{d/2-2}\ .
\label{sa1}
}

The structure of this result is very significant because if we
expand in powers of $k$ then a given term
\EQN{
k^{2n}\big(k_\mu
  k_\nu-k^2\delta_{\mu\nu}\big)
}
in momentum space corresponds to a term of the form
$F_{\mu\nu}\partial^{2n}F^{\mu\nu}$ in the effective action. 
This is because in momentum space
\SPN{
&F_{\mu\nu}(-k)F^{\mu\nu}(k)\longrightarrow\\ 
&-\big(ik_\mu A_\nu(-k)-ik_\nu A_\mu(-k)\big)
\big(ik^\mu A^\nu(k)-ik^\nu A^\mu(k)\big)
=-2\big(k_\mu k_\nu-k^2g_{\mu\nu}\big)A^\mu(-k)A^\nu(k)\ .
}
Since the
field strength $F_{\mu\nu}$ is gauge invariant,
the one-loop calculation is consequently entirely consistent with
gauge invariance. Taking the limit $\epsilon\to0$,
\eqref{sa1} includes the factor
\EQ{
\int_0^1dx\,(1-2x)^2\Big[\frac1{8\pi^2\epsilon}-\frac{\gamma_E}{16\pi^2}-\frac1{16\pi^2}\log\frac{x(1-x)k^2+m^2}{4\pi\mu^2}\Big]\ .
}
In the \MS\ scheme the required counter-term is
\EQN{
\LAG_\text{ct}=-\frac1{12}\Big[-\frac1{8\pi^2\epsilon}+\frac{\gamma_E-\log4\pi}
{16\pi^2}\Big]F_{\mu\nu}\,F^{\mu\nu}
}
where we used
\EQN{
\int_0^1(1-2x)^2=\frac13\ .
}
Being careful with the overall normalization, gives the one-loop effective 
action for the photon in Minkowski space in the \MS\ scheme:
\EQ{
S_\text{eff}[A_\mu]=-\frac14\int d^dk\,\Big[\frac1{e^2}-\frac1{16\pi^2}
\int_0^1dx\,(1-2x)^2\log\frac{m^2-x(1-x)k^2}{\mu^2}\Big]
F_{\mu\nu}(-k)F^{\mu\nu}(k)\ .
\label{hsa}
}
The one-loop beta function in the \MS\ scheme can then be deduced 
by demanding that $S_\text{eff}$ satisfies an RG equation of
the form \eqref{haq3} but with no wavefunction renormalization:
\EQ{
S_\text{eff}[A_\nu;\mu,e(\mu)]=S_\text{eff}[A_\nu;\mu',e(\mu')]\ .
}
This yields the beta function 
\EQ{
\mu\frac{de}{d\mu}=
\frac{e^3}{16\pi^2}\int_0^1dx\,(1-2x)^2
=\frac{e^3}{48\pi^2}\ .
\label{bfc}
}

It is interesting to repeat the calculation for a fermion. A Weyl
fermion$^{\ref{vp3}}$ couples to a gauge field via the kinetic term
\EQ{
\LAG_\text{kin}=i\bar\psi^{\dot\alpha}
\bar\sigma^\mu{}_{\dot\alpha\alpha}D_\mu\psi^\alpha\ .
}
In this case the fermion contributes to the vacuum polarization via the
diagrams 
\begin{center}
\includegraphics[height=0.7in]{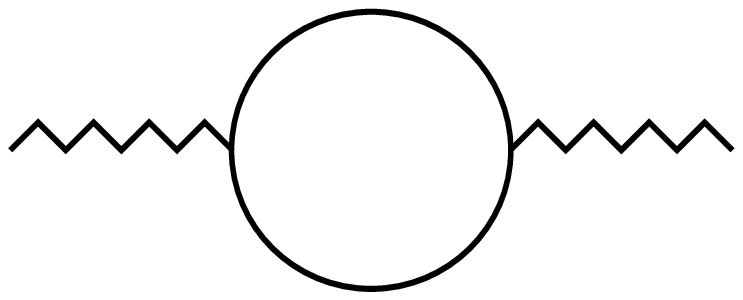}
\end{center}
If one follows through the derivation, the only difference in the
final result amounts to
replacement of $(1-2x)^2\to 4x(1-x)$ in \eqref{hsa}.$^{\ref{vp4}}$

Solving for the running coupling
\EQN{
\frac1{e(\mu)^2}=C-\frac1{24\pi^2}\log\mu\ ,~~~~\text{or}~~
e(\mu)^2=\frac{24\pi^2}{\log\Lambda/\mu}\ .
}
Notice that $e$, like the coupling $g_4$ in
scalar field theory, is irrelevant at the Gaussian fixed point. So $e(\mu)$
becomes smaller in the IR but has a Landau pole in the UV---at least
within the one-loop approximation. Hence QED, like scalar QFT, 
does not seem to have a continuum limit in $d=4$. Of course this does
not rule out a more complicated ``UV completion'' a issue that we will
return to later. 

There is tricky issue with regularization schemes like \MS\  which do
not involve setting conditions at a particular momentum, or energy,
scale. If we compare the beta functions in \eqref{sws},
and \eqref{gwg} then the former exhibit manifest decoupling meaning
that if we start
with some cut-off $\mu>m$ and flow down towards the IR
then when $\mu<m$ the RG flows 
are suppressed. This makes perfect physical sense: a
particle of mass $m$ only has effects when the energy scale is greater
than. However, in the \MS\ scheme \eqref{gwg} (or in the case of 
QED \eqref{bfc}) 
there does not appear to be any decoupling at all, the coupling is
still running when $\mu\ll m$. While there is nothing wrong with this in principle, it is
not very physical; in perturbation theory powers of
$e(\mu)\to0$ as $\mu\to0$ will be compensated
by powers of $\log\mu\to-\infty$ as in \eqref{hsa}.
Although, we shouldn't 
make the mistake of thinking of the couplings
in the action as physical couplings, we can introduce a more
physically motivated coupling in 
a momentum-independent regularization schemes like \MS\ by
decoupling the massive particle by hand when the RG scale $\mu$ crosses
the threshold at $\mu=m$. The recipe is as follows: two theories are written
down, one including the particle valid for $\mu\geq m$ and
one without the field for $\mu\leq m$. At $\mu=m$ physical quantities
in the the two theories are matched. For example, in 
scalar QED, for $\mu\geq m$ the coupling $e(\mu)$ 
runs as in \eqref{bfc}. At
$\mu=m$ the electron is removed from the theory to leave the
photon by itself. This is a non-interacting theory and the couplings does not
run, rather it is frozen at the value $e(m)$. The nasty $\log\mu$
factors disappear from perturbation theory to be left with a nice power
series in $e(m)$ which is a useful series 
if $e(m)$ is small enough, which in QED
is the case since $\tfrac{e(m_e)^2}{4\pi}=\tfrac1{137}$.

Hence, in this more
general philosophy as one flows into the IR, particles are decoupled
and removed from the theory as the RG scale $\mu$ passes the appropriate mass
thresholds. In a mass-dependent scheme decoupling is
automatic. For example, if instead of performing minimal subtraction 
we rather fix the value of the effective coupling in \eqref{hsa} at
the (Euclidean) momentum scale $k^2=\tilde\mu^2$ (see note \ref{mc1} of the
last lecture). This requires a counter-term
\EQN{
\LAG_\text{ct}=-\frac14\int_0^1dx\,(1-2x)^2\Big[
\frac1{8\pi^2\epsilon}-\frac1{16\pi^2}
\log\frac{x(1-x)\tilde\mu^2+m^2}{\mu^2}\Big]F_{\mu\nu}F^{\mu\nu}\ .
}
We now think of $\tilde\mu$ as the RG scale and 
the beta-function follows as
\EQN{
\tilde\mu\frac{de}{d\tilde\mu}=\frac{e^3}{16\pi^2}\int_0^1dx\,(1-2x)^2
\frac{\tilde\mu^2x(1-x)}{m^2+\tilde\mu^2x(1-x)}
}
which does indeed vanish when $\tilde\mu\ll m$---unlike \eqref{bfc}. A
comparison between the 2 running couplings in the form of
$e^{-3}\mu de/d\mu$ (the \MS\ one dashed) is shown below 
\begin{center}
\includegraphics[height=2.5in]{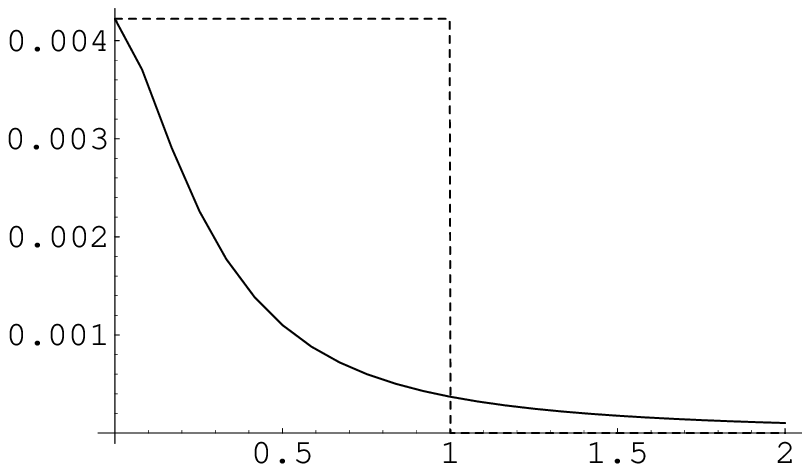}
\end{center}

\noindent{\bf\sl Non-abelian gauge theories}

Now we turn to the generalization of the RG flow of the gauge coupling
to non-abelian gauge theories. 
The Lagrangian for a non-abelian gauge field coupled to a massive scalar field
transforming in a representation $r$ of the gauge group $G$ has the form
\EQN{
{\cal
  L}=-\frac1{4g^2}
F_{\mu\nu}^aF^{a\,\mu\nu}+\big|D_\mu\phi_i\big|^2-m^2|\phi_i|^2\ ,
}
where
\EQN{
F_{\mu\nu}^a=\partial_\mu A^a_\nu-\partial_\nu A^a_\mu+f^{abc}
A^b_\mu A^c_\nu\ ,
}
and
\EQN{
D_\mu\phi_i=\partial_\mu\phi_i+iA^a_\mu T^a_{ij}\phi_j\ ,
}
where $f^{abc}$ are the structure constants of the gauge group $G$ and
$T^a_{ij}$ are the generators of a representation $r$ of $G$, so as matrices
\EQN{
[T^a,T^b]=if^{abc}T^c\ .
}
The RG flow of the gauge coupling $g$ can be determined in similar way to
QED. Integrating out the matter field produces a contribution which is
exactly \eqref{bfc} but multiplied by the group theory factor $C(r)$ where 
$C(r)\delta_{ab}=\text{Tr}\,T^aT^b$, for example, 
for gauge group $SU(N)$, for the adjoint representation $r=G$, $C(G)=N$ and
$C(\text{fund})=\tfrac12$. However, the new feature is that
there are contributions from the gauge field itself because of its
self-interactions. The relevant graphs are$^{\ref{vp5}}$
\begin{center}
\includegraphics[height=1in]{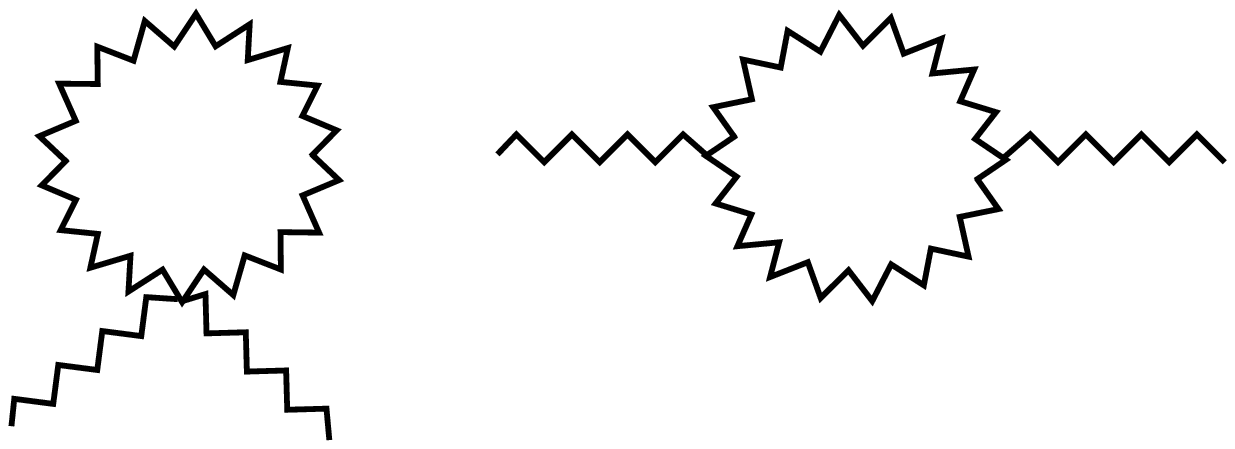}
\end{center}
The resulting expression for the one-loop beta-function including the
contributions from a series of scalar fields, as above, and, in
addition, Weyl fermions in representations $r_f$, all with a mass $m<\mu$, is
\EQ{
\mu\frac{dg}{d\mu}=
-\frac{g^3}{(4\pi)^2}\Big(\frac{11}3C(G)
-\frac13\sum_{\text{scalars }f}C(r_f)-\frac23
\sum_{\text{fermions }f}C(r_f)\Big)\ .
\label{bfd}
}
For example, for $G=SU(N)$ and $N_s$ scalars and
$N_f$ Weyl fermions in the fundamental $N$ representation plus the
anti-fundamental $\bar N$ 
\EQ{
\mu\frac{dg}{d\mu}=-\frac{g^3}{(4\pi)^2}\Big(\frac{11}3N
-\frac13N_s-\frac23N_f\Big)\ .
\label{laa}
}
While if all fields are in the adjoint representation
\EQ{
\mu\frac{dg}{d\mu}=-\frac{Ng^3}{(4\pi)^2}\Big(\frac{11}3-\frac13N_s-\frac23N_f\Big)\ .
\label{laa2}
}

What is remarkable is that the contribution from the self interactions
of the gauge field have an opposite sign from the matter fields. If we
write the one-loop beta function as $\mu dg/d\mu=-bg^3$ then the resulting
behaviour depends crucially on the sign of $b$. 
Solving the one-loop beta function equation $\mu dg/d\mu=-bg^3$ gives
\EQ{
g(\mu)^2=\frac1{C+b\log\mu}=\frac1{b\log(\mu/\Lambda)}\ ,
\label{fdd}
}
where we have written the integration constant as a dimensionful
parameter $\Lambda$. Once again we have an example of
{\it dimensional transmutation\/} where 
the dimensionless gauge coupling
has turned into a dimensionfull scale. 
When there is sufficient matter so that 
$b<0$, the situation will be as in QED: we must have $\mu<\Lambda$ and 
$g(\mu)$ will decrease with decreasing $\mu$, the theory is IR free, 
and so the coupling is marginally 
irrelevant at the Gaussian fixed point at $g=0$ and apparently no
continuum limit exists. In this case, $\Lambda$ is the
position of the Landau
pole in the UV as for $\phi^4$ theory.

On the contrary, as long as the amount of matter is not too large so
that $b>0$, we must have $\mu>\Lambda$ so that 
the coupling $g(\mu)$ increases with decreasing $\mu$. In this case
the coupling is marginally relevant at the Gaussian fixed
point and a non-trivial interacting continuum limit of the theory
exists since there is a renormalized trajectory that
flows out of the Gaussian fixed point. The situation is rather
different from scalar fields in $d<4$ because the UV limit of the
theory is non-interacting, a situation known as asymptotic freedom.
In this case, $\Lambda$ signals the scale at which the coupling 
becomes large and perturbation theory breaks down. It is at these
scales that confinement sets in. It is important to realize that
$\Lambda$ is a physically measurable scale.

\noindent{\bf\sl The Higgs mechanism}

There is another mechanism which affects the running of the gauge
coupling. This is the phenomenon of {\it spontaneous symmetry
  breaking\/} or the {\it Higgs effect\/} which is driven by
VEVs for the scalar fields. In general, a VEV $\phi^0$ will not be
invariant under gauge transformations. In other words, we can split
the generators of the gauge group into 2 sets ${\cal H}\cup{\cal
  B}$:
\EQ{
T^a_{ij}\phi^0_j=0\qquad a\in{\cal H}\ ,~~~~T^a_{ij}\phi^0_j
\neq0\qquad a\in{\cal B}\ .
}
The first set are the unbroken symmetries which generate a subgroup
$H\subset G$. The second set are the broken symmetries. The gauge
fields corresponding to the broken symmetries gain a mass which is
determined by the kinetic term of the scalar:
\EQ{
D_\mu\phi^\dagger D^\mu\phi\longrightarrow 
\sum_{a,b\in{\cal B}}\Big(\phi^{0\dagger}T^a
T^b\phi^0\Big)\, A^a_\mu A^{b\mu}\ .
}

For example, consider a scalar field in the $\bN$ of $SU(N)$. Up to
gauge transformation we can suppose the VEV takes the form
\EQ{
\phi^0_i=v\delta_{i1}\ .
}
If we represent the gauge field as a traceless $N\times N$ matrix,
then 
\EQ{
A^a_\mu T^a_{ij}\phi_j^0=A_{\mu ij}\phi_j^0=v A_{\mu i1}\ .
}
Hence, the components $A_{\mu i1}=A_{\mu 1i}^*$ gain a mass 
$v$ and the gauge group is broken from $SU(N)$ to
$SU(N-1)$. It should not be surprising that at the scale $\mu=v$
the beta function of the gauge coupling changes discontinuously (in a
momentum independent scheme like \MS) from that appropriate to a gauge
group $G$ to one associated to a gauge group $H$.

\BOX{\begin{center}{\bf UV Completion}\end{center}

We have seen that an essential requirement in order to take a continuum
limit in a QFT is that the theory in the IR lies on a
renormalized trajectory, that is the RG flow out of some UV fixed
point. For two theories that we have met, 
$\phi^4$ scalar QFT and QED, both in $d=4$, there is no renormalized
trajectory issuing from a fixed point and so
when we try to take the continuum limit 
there is a Landau pole: 
as we run the RG backwards the
coupling diverges to $\infty$ at some finite $\mu$.\\

If this was the last word on continuum limits, then this would 
seem to be rather disastrous for the
standard model which includes QED and the Higgs scalars. However, we
have already seen that as the RG scale passes a mass of particle into
the regime $\mu<m$, the particle is essentially redundant as far as
the physics is concerned and in RG
schemes like \MS\ is actually removed by hand from the theory
(although it is important to remember that it leaves its mark in the
form of irrelevant operators suppressed by powers of 
$1/m$ which in principle will leave observable effects.)
So in our theories
with Landau poles, it may be that there is a good {\it UV
  completion\/} which requires new particles and fields as we flow
into the UV. For example, QED, along with the other gauge interactions
of the standard model with group $U(1)\times SU(2)\times SU(3)$ 
can be embedded in a larger group, the simplest being $SU(5)$. This is
idea of grand unification. The $SU(5)$ is then broken to the standard
model, including QED, 
by certain Higgs fields at certain energy scales such that the
standard model is the effective theory at low energies. However, in the
UV the theory is asymptotically free and so there exists a continuum
limit (at least if we ignore the Landau pole in the Higgs sector). 
The Higgs sector of the standard model can also be given a consistent
UV completion in models where it is the bound state of fermions, for
example.
}

\begin{center}{\bf Notes}\end{center}

\begin{quote}
{\small

\begin{note}Of course this will only be true if the cut-off scheme
  itself does not break gauge invariance. For instance the sharp
  momentum cut-off is not consistent with gauge invariance because
  gauge transformations mix up modes of different frequency, {\it
  i.e.\/}~they do not preserve the split \eqref{dcp}.
\label{vp1}
\end{note} 

\begin{note}A Weyl fermion $\psi_\alpha$, $\alpha=1,2$ and
    $\aD=1,2$, is a
    2-component complex Grassmann quantity. In Minkowski space 
its conjugate is denoted
    $\bar\psi^\aD=(\psi_\alpha)^\dagger$. In Euclidean space
    $\psi_\alpha$ and $\bar\psi^\aD$ are treated as independent real
    objects. A Weyl fermion can also be represented in 4-component
    Dirac fermion language as 
 a   Majorana fermion $\Psi=\MAT{\psi_\alpha\\ \bar\psi^\aD}$. A
    Dirac fermion, like the electron-positron field 
is then made up of two Weyl fermions $\Psi=\MAT{\psi_\alpha\\ 
\bar\lambda^\aD}$. In this 2-component notation the $\sigma$ matrices
    can be taken as
\EQN{
\sigma^\mu_{\alpha\aD}=\big({\bf 1},\sigma^i\big)\ ,\qquad
\bar\sigma^{\mu\aD\alpha}=\big({\bf 1},-\sigma^i\big)\ .
}
\label{vp3}\end{note}

\begin{note}In general if we have a set of complex scalars and Weyl
    fermions with integer charges $q_i$ meaning that covariant
    derivatives involve $D_\mu=\partial_\mu+ieq_iA_\mu$, or
    $\partial_\mu+iq_iA_\mu$ after the rescaling $A_\mu\to A_\mu/e$, 
then the one-loop beta function in the \MS\ scheme is
\EQ{
\mu\frac{de}{d\mu}=\frac1{48\pi^2}\sum_\text{scalars} q_i^2+\frac1{12\pi^2}
\sum_\text{fermions}q_i^2\ .
}
\label{vp4}\end{note}

\begin{note}Depending on how the gauge is fixed, there will usually be
    a contribution from the gauge-fixing ghost fields. The complete
    calculation can be found in Peskin and Schroeder p533 onwards.
\label{vp5}\end{note}

}\end{quote}

\newpage
\section{RG and Supersymmetry}

QFTs with supersymmetry (SUSY) have some remarkable RG properties.
From our point-of-view, one can prove the existence of many non-trivial fixed
points of the RG. Our discussion of SUSY theories will be geared
towards describing some of these fixed points and the associated RG
flows and so will be very brief about other aspects of SUSY
theories. In particular, 
we shall use component fields rather than introduce the whole paraphernalia of
superspace and superfields and will be rather brief when discussing
the global R-symmetries of SUSY theories.

In a SUSY theory, fields are collected into SUSY
multiplets. In $d=4$ (which we will stick to from now on) 
there are 2 basic multiplets: (i) a chiral multiplet
consisting of a complex scalar, a Weyl fermion and a complex auxiliary
scalar field
\EQN{
\Phi=(\phi,\psi_\alpha,F)\ ,
}
(ii) a vector multiplet consisting of a gauge field, an adjoint-valued
Weyl fermion and an adjoint-valued auxiliary real scalar field 
\EQN{
V=(A_\mu,\lambda_\alpha,D)\ .
}

\noindent{\bf\sl Theories of chiral multiplets: Wess-Zumino models}

Wess-Zumino models are constructed from chiral multiplets $\Phi_i$ 
with a basic SUSY kinetic term of the form$^{\ref{s4}}$
\EQN{
\LAG_\text{kin}=
|\partial_\mu\phi_i|^2+i\bar\psi_i\bar\sigma^\mu\partial_\mu
\psi_i+|F_i|^2\ ,
}
note that the auxiliary fields $F_i$ has trivial kinetic terms and their
r\^ole in the theory is to simplify the SUSY structure. 
The
interactions are determined by a function $W(\phi_i)$, the superpotential,
\EQN{
\LAG_\text{int}=F_i\frac{\partial W}{\partial\phi_i}+
\frac{\partial^2W}{\partial\phi_i\partial\phi_j}\psi_i\psi_j+\text{c.c.}
}
(c.c.$=$complex conjugate.)
Notice that the $F_i$ can trivially be ``integrated out'' by their 
equation-of-motion
\EQN{
F_i=\frac{\partial W(\phi^*)}{\partial\phi_i^*}\ ,
}
to give a net potential on the scalar fields:
\EQ{
V(\phi_i,\phi_i^*)=\sum_i\left|\frac{\partial W}{\partial\phi_i}\right|^2\ .
\label{spo}
}
The theory is invariant under the infinitesimal SUSY
transformations Grassmann parameter $\xi_\alpha$ 
\SP{
\delta\phi_i&=\xi\psi_i\ ,\\
\delta\psi_i&=i\sigma^\mu\bar\xi\partial_\mu\phi_i+\xi F_i\ ,\\
\delta F_i&=i\bar\xi\bar\sigma^\mu\partial_\mu\psi_i\ .
\label{str}
}

There is an important distinction between the kinetic and interaction
terms: $\LAG_\text{kin}$ are 
``D terms'', meaning functions of the fields and
their complex conjugates, while $\LAG_\text{int}$ are ``F terms'',
consisting
of the sum of a part which is holomorphic in the fields and the couplings
and the complex conjugate term which has all 
quantities replaced by the corresponding anti-holomorphic
ones. In particular, the superpotential $W(\phi_i)$ is
a holomorphic function of the $\phi_i$ and also the couplings $g_n$, {\it
  i.e.\/}~$W(\phi_i)$ does not depend on $\phi^*_i$ or $g_n^*$. 
It is this underlying holomorphic structure of F-terms 
that plays the pivotal r\^ole in proving
exact RG statements in SUSY theories.

For example, for a $\phi^4$ type model the superpotential has the form
\EQ{
W(\phi)=\frac12m\phi^2+\frac13\lambda\phi^3\ ,
\label{mcc}
}
and note that the coupling $m$ and $\lambda$ are in general complex. 
In this case, after integrating out the auxiliary fields $F_i$, the
interaction terms are in this case
\EQN{
{\cal
  L}_\text{int}=-\big|m\phi+\lambda\phi^2\big|^2-m\psi\psi-m^*\bar\psi\bar\psi
-2\lambda\phi\psi\psi-2\lambda\phi^*\bar\psi\bar\psi\ .
}
The final 2 terms here are Yukawa interactions between the fermions
and scalars.

\noindent{\bf\sl Renormalization of the F-terms}

From the point-of-view of RG, the key fact 
is that couplings in the superpotential
do not change when we change the RG scale $\mu$.$^{\ref{s3}}$ This does
{\it not\/} mean that the couplings in the superpotential do not have
any RG flow because there can be be non-trivial wavefunction
renormalization, in addition to the
canonical scaling coming from the fact that we always think of RG flow of
dimensionless couplings. In fact,
for a term $W\sim 
\mu^{3-n}\lambda\phi^n$, the RG equation becomes
\EQN{
W(Z(\mu)^{1/2}\phi;\mu,\lambda(\mu))
=W(Z(\mu')^{1/2}\phi;\mu',\lambda(\mu'))
}
and so under an RG flow
\EQN{
\mu^{3-n}Z(\mu)^{-n/2}\lambda(\mu)=\mu^{\prime{3-n}}
Z(\mu')^{-n/2}\lambda(\mu')\ ,
}
from which we extract the beta-function$^{\ref{s1}}$
\EQN{
\mu\frac{d\lambda}{d\mu}=(-3+n+n\gamma)\lambda\ ,
}
where $\gamma$ is the common anomalous dimension of all fields in the chiral
multiplet (SUSY ensures that all fields in a multiplet have the same
wavefunction renormalization). 
Another way to state this, is that the scaling dimension of the
{\it composite operator\/} $\phi^n$ is just equal to the sum of the scaling
dimension of the individual operators $\phi$:$^{\ref{s2}}$
\EQN{
\Delta_{\phi^n}=n(1+\gamma)=n\Delta_\phi\ .
}
On the contrary, SUSY does {\it not\/}
constraint the running of D-term couplings.

The fact that the superpotential is only renormalized by wavefunction
renormalization has a very important application. The vacua of a
theory are determined by minimizing the effective
potential. This latter quantity is, as argued previously, 
the potential in the
Wilsonian effective action in the limit that $\mu\to0$.
In a SUSY theory, the potential, is given by \eqref{spo} and so the absolute
minima of $V$ correspond to the critical points of the superpotential,
or ``F-flatness'' condition
\EQ{
F_i^*=-\frac{\partial W(\phi_j)}{\partial\phi_i}=0\ .
\label{kdd}
}
Such vacua can be shown to preserve SUSY, whereas if the minima have
$V>0$ then SUSY is spontaneously broken. Now we come to the important bit,
as we decrease the cut-off 
$\mu\to0$, if we have a solution $\phi_i$ of \eqref{kdd}
at the original cut-off, then we still have a solution,
differing only by wavefunction renormalization, $Z_i(\mu)^{1/2}\phi_i$.

For example, the model with a bare superpotential \eqref{mcc} has SUSY
vacua at $\phi=0$ and $\phi=-m/\lambda$ and by the argument above
these vacua will persist, only shifted by wavefunction
renormalization, once the quantum corrections are taken into
account. It is also possible to cook-up models which do not have any
SUSY vacua.

\BOX{\begin{center}{\bf Vacuum Moduli Spaces}\end{center}

In a non-SUSY QFT, vacua of the theory are generally discrete points
because RG flow generically lifts any flat directions of
the potential. In a SUSY theory, however, RG flow cannot 
lift a degenerate space of SUSY vacua. So in many cases SUSY theories
have non-trivial Vacuum Moduli Spaces ${\mathfrak M}$ whose points are
labelled by the VEVs of the scalar fields in the theory.  The space
${\mathfrak M}$ is in many cases not a manifold rather it is a series
of manifolds, or ``branches'', often joined along subspaces of lower
dimension.\\

As an example consider a model with 3 chiral multiplets
$X$, $Y$ and $Z$ with a superpotential $W=\lambda xyz$. The SUSY vacua 
are determined by the equations $xy=yz=zx=0$ and so there are 3
branches (i) $y=z=0$, $x$ arbitrary, (ii) $z=x=0$, $y$ arbitrary, and
(iii) $x=y=0$, $z$ arbitrary. Here, $x$, {\it etc\/}, is short for the
VEV of the scalar field component of $X$. 
These three branches are joined at the point $x=y=z=0$. The structure
of ${\mathfrak M}$ is schematically
\begin{center}
\includegraphics[width=1.8in]{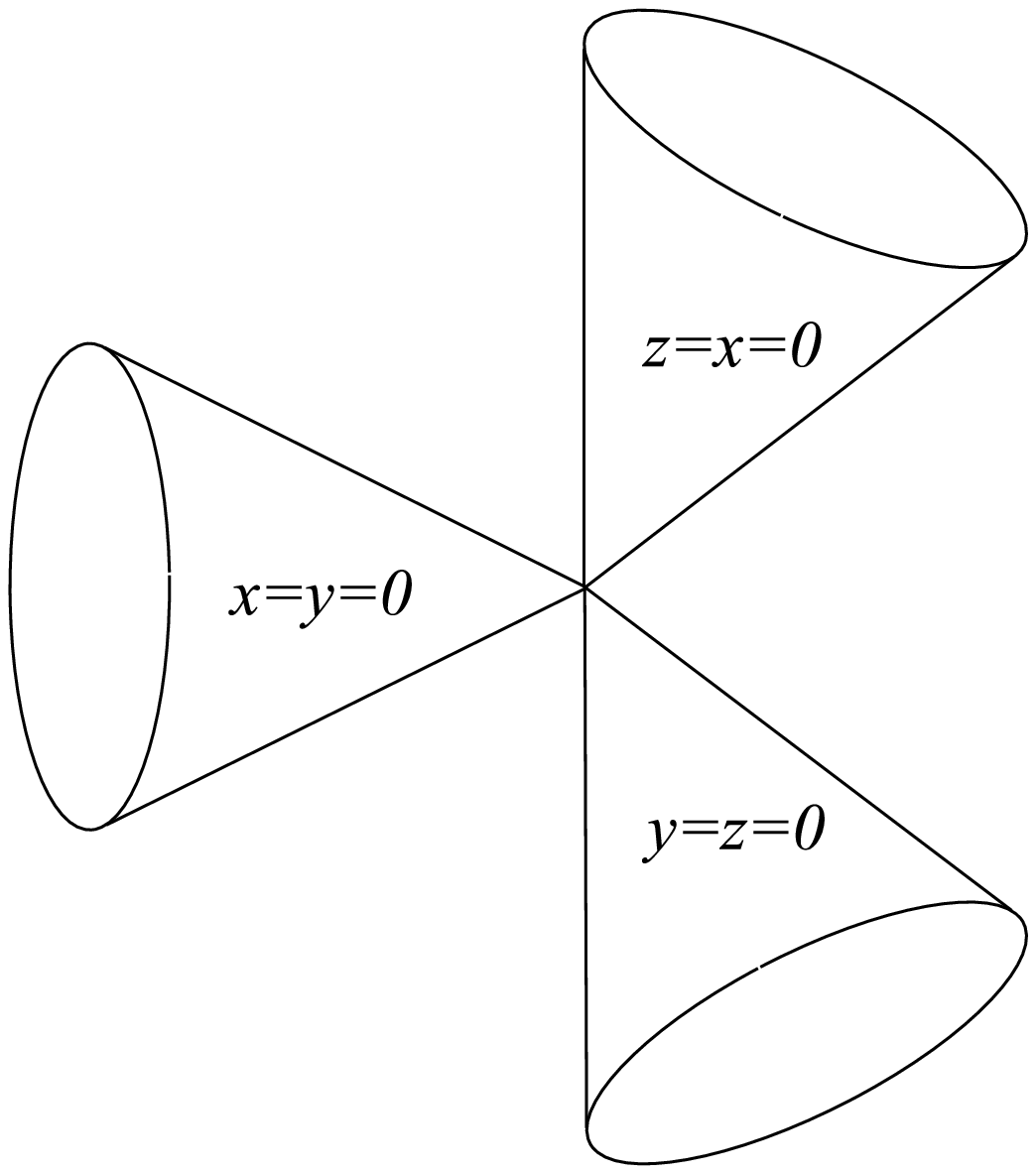}
\end{center}
}

\noindent{\bf\sl SUSY gauge theories} 

A SUSY gauge theory is constructed from a kinetic term which, unlike
the chiral multiplet, is an F-term
\SP{
\LAG_\text{kin}(V)&=
\Big(\frac1{2g^2}+\frac{\theta}{16\pi^2 i}\Big)
\LAG(V)+\text{c.c.}\ ,\\
\LAG(V)&=
-\frac14F^{a}_{\mu\nu}F^{a\mu\nu}+\frac i{8}\epsilon_{\mu\nu\rho\sigma}
F^{a\mu\nu}F^{a\rho\sigma}-i\bar\lambda^a\bar\sigma^\mu
D_\mu\lambda^a+\frac12D^{a2}\ .
\label{xax}
}
Notice that the gauge
coupling is naturally combined with the $\theta$ angle$^{\ref{s5}}$ to form a
holomorphic coupling which is conventionally written as 
\EQN{
\tau=\frac{4\pi i}{g^2}+\frac{\theta}{2\pi}\ .
}

If we couple a vector superfield to a chiral superfield transforming
in some representation $r$, with generators $T^a_{ij}$, of the gauge group, 
then SUSY requires minimal
coupling: all derivatives are replaced by covariant derivatives. In
addition, there are the extra interactions
\EQN{
\LAG_\text{int}=i\sqrt2
T^a_{ij}\big(\phi^*_i\psi_j\lambda^a-\phi_i\bar\psi_j\bar\lambda^a\big)
+D^aT^a_{ij}\phi_i^*\phi_j\ .
}
Notice that once the auxiliary fields $D^a$ and $F_i$ are 
integrated-out, the scalar fields have a potential
\EQ{
V(\phi_i,\phi^*_i)=\left|\frac{\partial W}{\partial\phi_i}\right|^2
+\frac12\sum_a\big(\phi_i^*T^a_{ij}\phi_j\big)^2\ ,
\label{sgp}
}
which generalizes \eqref{spo}. Henceforth, we will think of $\phi$ as
a vector and leave the indices implicit. If we have several matter
fields transforming in representations $r_f$ 
then we will denote them with a ``flavour index'' $\Phi_f$. For
instance, SUSY QCD with gauge group $SU(N)$ is defined as the theory
with $N_f$ chiral multiplets in the $\bN$ representation,
conventionally denoted $Q_f=(q_f,\psi_f,F_f)$, and  
$N_f$ chiral multiplets in the $\bar\bN$ representation,
denoted $\tilde Q_f=(\tilde q_f,\tilde\psi_f,\tilde F_f)$.

Since the gauge coupling is a secretly a holomorphic quantity one
wonders whether it has any non-trivial renormalization. The reason is
that the perturbative expansion is in $g$ and not $\tau$, but
renormalization must preserve the holomorphic structure hence there
should be no RG flow of $\tau$ and hence $g$. Actually, this argument
is a bit too quick, because one-loop running is consistent with holomorphy,
since for the theory without matter fields, using \eqref{bfd} with one
fermion in the adjoint representation, we have
\EQN{
\mu\frac{dg}{d\mu}=-\frac{3}{16\pi^2}C(G) g^3~~~~~\implies~~~~~
\mu\frac{d\tau}{d\mu}=\frac{3i}{2\pi}C(G)\ .
}
which is consistent because $\theta$ does not run. So the beta function
of $g$ is exact at the one-loop level!

\noindent{\bf\sl The re-scaling anomaly}

Once we add chiral multiplets coupled to the vector
multiplet the beta function of $g$ will get non-trivial
contributions for two reasons. The first effect is simple: the one-loop 
coefficient receives additional contributions from the fields of the chiral
multiplets. From \eqref{bfd}, and taking account the field content of the
vector and chiral multiplets, we simply have to perform the replacement
\EQN{
3C(G)\longrightarrow 3C(G)-\sum_{f}C(r_f)\ .
}
The second contribution is more subtle. As we have learned
under the RG transformation, as the cut-off $\mu$ is lowered 
the kinetic terms changes by the wavefunction renormalization
factor:
\EQN{
|\partial_\mu\phi|^2
\longrightarrow Z|\partial_\mu\phi|^2\ .
}
At this point, we have to perform the re-scaling
$\phi\to Z^{-1/2}\phi$ in order to return the kinetic term to its
canonical form. In principle, when we perform this re-scaling we 
should take into account the Jacobian coming from the measure of the
functional integral. With a sharp
momentum cut-off on the modes this gives a Jacobian
\EQN{
\text{Jac}=Z^\text{\# degrees-of-freedom}=\exp\Big[\log\,Z\,\int d^dx\,
\int_{|p|\leq\mu}\frac{d^dp}{(2\pi)^d}\Big]
=\exp\Big[\log Z\,\mu^d\,\frac{\text{Vol}\,S^{d-1}\int d^dx}{(2\pi)^d}\Big]\ .
}
This is divergent but still just a constant factor that we can safely ignore.
However, when $\phi$ is coupled non-trivially
to a gauge field, the Jacobian will depend on the 
background gauge field. The reason is in order to preserve gauge
invariance the cut-off must somehow be compatible with the covariant
derivative $D_\mu=\partial_\mu+iA_\mu$ 
and this means that the whole cut-off procedure must involve
the background gauge field in a non-trivial way. The proper way to do
this is to put the cut-off $\mu^2$ on the eigenvalues of the covariant
Laplace equation (in Euclidean space)
\EQN{
-D_\mu^2\phi=\lambda\phi\ .
}
It is technically quite complicated to calculate the Jacobian this
way; however, we can avoid this by making an 
intelligent use of the chiral anomaly.

\BOX{\begin{center}{\bf The Chiral Anomaly}\end{center}

A chiral transformation takes
\EQN{
\psi\to e^{i\alpha}\psi\ ,\qquad
\bar\psi\to e^{-i\alpha}\bar\psi\ .
}
The action is clearly invariant under this transformation, however, 
the regularized measure $[d\psi][d\bar\psi]$ is not. 
The point is that the cut-off procedure breaks the symmetry in the
presence of a background gauge field. Under the
transformation the fermion part of the measure picks up a non-trivial
Jacobian 
\EQ{
[e^{i\alpha}d\psi][e^{-i\alpha}d\bar\psi]=
[d\psi][d\bar\psi]\exp\Big[ -
\frac{i\alpha C(r)}{32\pi^2}
\int d^4x\,\epsilon_{\mu\nu\rho\sigma}
F^{a\mu\nu}F^{a\rho\sigma}\Big]\ .
\label{zaz}
}
What is remarkable about the chiral anomaly is that this
result is exact to all orders in perturbation theory. 
}

We can now use the chiral anomaly to calculate the re-scaling anomaly,
or Jacobian, for a
chiral multiplet by exploiting holomorphy and SUSY. If one compares
the 
right-hand side of \eqref{zaz} with the SUSY kinetic term \eqref{xax},
taking into account that the holomorphic and anti-holomorphic
contributions to $\LAG_\text{kin}$ are separately invariant under SUSY,
it is clear that \eqref{zaz} can be written in a way manifestly
invariant under SUSY; namely 
\EQ{
\text{Jac}=
\exp\Big\{\frac{\alpha C(r)}{8\pi^2}\int d^4x\,\Big[\LAG(V)-
\LAG^*(V)\Big]\Big\}
=\exp\Big\{-\frac{iC(r)}{8\pi^2}\int d^4x\,
\Big[\big(i\alpha\LAG(V)\big)+\big(i\alpha\LAG(V)\big)^*\Big]\Big\}\ .
\label{sms}
}
Most of the terms here cancel to leave only the right-hand side of 
\eqref{zaz}. However, we can now exploit holomorphy to 
generalize the chiral transformation to
an arbitrary holomorphic re-scaling of
the whole chiral multiplet:
\EQN{
\phi\to Z^{-1/2}\phi\ ,\qquad \psi\to Z^{-1/2}\psi\ ,\qquad 
F\to Z^{-1/2}F\ .
}
The resulting Jacobian is then simply the right-hand side of \eqref{sms} with
$e^{i\alpha}$ replaced by $Z^{-1/2}$:
\EQN{
\text{Jac}=\exp\Big\{ \frac{iC(r)}{16\pi^2}
\int d^4x\,\Big[\big(\log Z\,\LAG(V)\big)+
\big(\log Z\,\LAG(V)\big)^*\Big]\Big\}\ .
}

Now this result 
is also valid, by analytic continuation, 
for a field wavefunction renormalization for which $Z$ is real.
Hence, we can calculate the effect of wavefunction renormalization 
on the running of the gauge coupling. Under an
infinitesimal RG transformation $Z=1-2\gamma\delta\mu/\mu$ and so the 
Jacobian is of the form
\EQN{
\text{Jac}=\exp\Big\{ -\frac{iC(r)\gamma}{8\pi^2}
\int d^4x\,\big[\LAG(V)+\LAG^*(V)\big]\frac{\delta\mu}\mu\Big\}
}
which corresponds to an additional contribution to the flow of the
gauge coupling of 
\SPN{
\delta\Big(\frac1{g^2}\Big)&=\frac{C(r)\gamma}{4\pi^2}\cdot\frac{\delta\mu}\mu
\ ,\\ 
\mu\frac{dg}{d\mu}\Big|_\text{additional}&=-\frac{g^3}{8\pi^2}C(r)\gamma\ .
}
Hence, the exact beta function for a model with a series of  
chiral multiplets in representations $r_f$ of the gauge group 
is
\EQ{
\mu\frac{dg}{d\mu}
=-\frac{g^3}{16\pi^2}\Big(3C(G)-\sum_fC(r_f)(1-2\gamma_f)\Big)\ .
\label{hh1}
}

\noindent{\bf\sl The NSVZ exact beta-function}

The above is an exact result valid for the coupling that appears as $1/g^2$ 
in front of the gauge kinetic term. We would think that this is the
same as the {\it canonical gauge coupling\/} $g_c$, the one
that appears in the covariant derivatives
$D_\mu=\partial_\mu+ig_cA_\mu$ because one can simply re-scale the
gauge field $A_\mu\to g_c A_\mu$ (and all the
other fields of the vector multiplet). But we have just learnt our
lesson: these re-scalings cannot be done willy-nilly since 
one can expect 
a non-trivial Jacobian from measure and this means that $g\neq g_c$. 
Unfortunately the trick that
worked for a chiral multiplet does not work in a simple way
here since the chiral
rotation of the gluino cannot be complexified to give the re-scaling
anomaly of the vector multiplet, and so we just quote the result
\EQN{
\text{Jac}
=\exp\Big\{ \frac{iC(G)g^2}{16\pi^2}\log g_c
\int d^4x\,\Big(\LAG(V)+\LAG(V)^*\Big)\,\frac{\delta\mu}\mu\Big\}\ .
}
Notice that it has the same form as an adjoint-valued chiral multiplet
but with the opposite sign and with 
$Z=g_c^2$. The condition that the kinetic term, after
the re-scaling, has no coupling in front of the kinetic term 
gives the condition
\EQN{
g_c^2\Big(\frac1{g^2}-\frac{C(G)}{4\pi^2}\log g_c\Big)=1
}
which is the exact relation between the two definitions of the gauge
coupling $g$ and $g_c$. 
It follows that the beta function of the canonical coupling
is
\SP{
\mu\frac{dg_c}{d\mu}&=\mu\frac{dg_c}{d\mu}=\mu\frac{dg}{d\mu}\cdot\frac{g_c^3}
{g^3}\cdot\frac1{
1-C(G)g_c^2/8\pi^2}
\\ &=-\frac{g_c^3}{16\pi^2}\cdot\frac{3C(G)-\sum_fC(r_f)(1-2\gamma_f)}{
1-C(G)g_c^2/8\pi^2}\ .
\label{hh2}
}
This is the famous Novikov-Shifman-Vainshtein-Zahakarov (NSVZ) beta
function for SUSY gauge theories.

\noindent{\bf\sl Vacuum structure}

Searching for the vacua of a SUSY gauge theory is more
complicated than for a Wess-Zumino model. 
The potential of a SUSY gauge theory \eqref{sgp} includes a
contribution from the gauge sector which arises once the $D$ field is
integrated out. Clearly, as in the Wess-Zumino case, it is
bounded below by 0 and for a SUSY vacuum we need $V=0$ and so the conditions
are
\SP{
\frac{\partial W}{\partial\phi_f}=0\ ,\\
\sum_f \phi^\dagger_fT^a\phi_f=0\ ,
}
known as the F- and D-term equations, 
which have to solved moduli gauge transformations. Such minima correspond
to vacua of the theory which preserve SUSY.

To see how this works, consider SUSY QED with chiral
fields $Q_1$ and $Q_2$ of charge $+1$ and $\tilde Q_1$ and $\tilde
Q_2$ of charge $-1$. Furthermore, let us suppose that the 
theory has no superpotential. Hence, only the D-term
equation is non-trivial:
\EQN{
|q_1|^2+|q_2|^2-|\tilde q_1|^2-|\tilde q_2|^2
=0\ .
\label{haa}
}
One way to solve this is to 
use gauge transformations to make $q_1$ real and then
the D-term equation determines, say, 
$q_1$. This breaks down when $q_1=0$, but then we can use $q_2$ instead, {\it
  etc.\/} All-in-all, modulo gauge transformation, 
we have a 3-complex dimensional space of
solutions which defines the vacuum moduli space 
${\mathfrak M}$. At any
point in ${\mathfrak M}$, besides the origin, the VEV of the scalar fields
breaks the gauge group and the photon gains a mass by the Higgs
mechanism.

It is an important theorem, that any solution of the F- and D-term 
equations moduli gauge transformations is
equivalent to a solution of the F-term equations {\it alone\/} modulo {\it
  complexified\/} gauge transformations.$^{\ref{s9}}$ This theorem is
useful because it survives quantum corrections. In particular, 
although the D-terms are renormalized non-trivially, we can ignore
them if we are only interested in the SUSY vacua of theory. In the
present case, complexifed gauge transformations acts as
\EQN{
q_1\to \lambda q_1\ ,\qquad q_2\to \lambda q_2\ ,\qquad
\tilde q_1\to \lambda^{-1}\tilde q_1\ ,\qquad \tilde q_2\to 
\lambda^{-1}\tilde q_2\ .
}

So for the example above, which has no F-term conditions, we can find
${\mathfrak M}$ by finding a set if coordinates which are gauge
invariant in a complexified sense: in this
case $z_1=q_1\tilde q_1$, $z_2=q_1\tilde q_2$, $z_3=q_2\tilde q_1$ and
$z_4=q_2\tilde q_2$. However, these coordinates are not all
independent, a short-coming that can be remedied by imposing the condition
\EQN{
z_1z_4=z_2z_3\ .
}
So the vacuum moduli space is an example of a 
{\it complex variety\/} defined by
the above condition in ${\bf C}^4$. In this case the complex
variety is actually a {\it conifold\/} rather than a manifold since it
is singular at $z_i=0$ where the $U(1)$ gauge symmetry is restored. 

\begin{center}{\bf Notes}\end{center}

\begin{quote}
{\small

\begin{note}The kinetic term can be generalized to all the terms with 
  two derivatives:
\EQ{
\LAG_\text{kin}=\frac{\partial^2K}{\partial\phi_i\partial\phi_j^*}
\partial_\mu\phi_i\partial^\mu\phi_j^*+\text{terms involving
}(\psi,F)\ ,
}
determined by a real function,
$K(\phi,\phi^*)$, the {\it K\"ahler potential}.
\label{s4}
\end{note}

\begin{note}The simple proof of this fact is to think of the
  holomorphic and anti-holomorphic fields and couplings as being
  temporarily independent. Then when we integrate out modes as we
  decrease $\mu$ the flow of the holomorphic couplings cannot depend
  on the anti-holomorphic couplings and so we can set the latter to
  0. In this limit, both the scalar and fermion propagator have no
  components which can joint the legs of holomorphic fields. Hence,
  the graphs that contribute to the holomorphic fields simply vanish.   
\label{s3}
\end{note}

\begin{note}The 3 here is the canonical dimension of $W$ and
  $\gamma$ is the anomalous dimension of each field of the chiral
  multiplet which must be equal by SUSY. The generalization for a term
  $W\sim \mu^{3-p}\lambda\phi_{i_1}\cdots \phi_{i_n}$ is
\EQ{
\mu\frac{d\lambda}{d\mu}=\big(-3+p+\sum_{i=1}^n\gamma_i\big)\lambda\ .
\label{uu1}
}
\label{s1}
\end{note}

\begin{note}It is a key aspect of RG that scaling dimension of a
  general composite operator ${\cal O}={\cal O}_1\cdots{\cal O}_p$ is
  {\it not\/} usually the sum $\Delta_{\cal O}\neq\Delta_{{\cal
  O}_1}+\cdots+\Delta_{{\cal O}_p}$. For instance, for $\phi^4$ theory
  around the Wilson-Fischer fixed point
  $\Delta_{\phi^4}\neq2\Delta_{\phi^2}$. 
\label{s2}
\end{note}

\begin{note} The $\theta$ angle multiplies a term
\EQN{
\frac1{64\pi^2}\int d^4x\,\epsilon_{\mu\nu\rho\sigma}
F^{a\mu\nu}F^{a\rho\sigma}\ ,
}
in the action. This integral is topological, in the sense that for any
smooth gauge configuration it is equal to $2\pi k$, where $k$ is an
integer: the $2^\text{nd}$ Chern Class of the gauge field. 
Furthermore, the theta term does not
contribute to the classical equations-of-motion. In the quantum theory
which involves the Feynman sum over configurations, $\theta$ is another
coupling in the theory.   
\label{s5}
\end{note}

\begin{note}We present the proof in the abelian case $G=U(1)$. 
Suppose we have a solution of the
  F-term equations $q_f$ (with charges $e_f$). The key point is that
  since the superpotential only depends on the holomorphic fields and
  not the anti-holomorphic ones, it is actually invariant not just
  under gauge transformations $q_f\to e^{ie_f\theta}q_f$ but also
  complexified gauge transformations $q_f\to \eta^{e_f}q_f$, for
  arbitrary complex $\eta$. On the other hand, the D-term 
\EQN{
\sum_f e_f|q_f|^2
}
is not invariant under this complexified transformation. The strategy
  is then to use the complexified gauge transformation with $\eta$
  real, to find a
  solution of the D-term equation: $\check{q}_f=\eta^{e_f}q_f$:
\EQN{
\sum_fe_f|\check q_f|^2=\sum_f
  e_f\eta^{2e_f}|q_f|^2=\frac12\frac{\partial}{\partial\eta}\sum_f
 \eta^{2e_f}|q_f|^2=0\ .
}
It is always possible to find an $\eta$ which does this because for the
  function $\sum_f  \eta^{2e_f}|q_f|^2$ either (i) $e_f$ all have the
  same sign, in which case $\eta\to0$ or $\infty$ depending on the
  sign of the charges, or (ii)
 if some of the $e_f$ have opposite sign 
$\sum_f  \eta^{2e_f}|q_f|^2$ goes to $\infty$
  for $\eta\to0,\infty$, and hence has a minimum as a function of
  $\eta$ for some finite $\eta>0$.

The generalization to the non-abelian case is reasonably
straightforward: see Wess and Bagger p.57-58.
\label{s9}
\end{note}

}\end{quote}

\newpage
\section{More RG Structure of SUSY Theories and $\N=4$}

\noindent{\bf\sl RG fixed points}

The next issue we will address 
in our RG investigation of SUSY theories will be to show that there
are many non-trivial fixed points of the RG in $d=4$ supersymmetric
gauge theories. These fixed point have a supersymmetric extension of 
conformal invariance and are consequently 
super-conformal field theories (SCFTs). The existence of these fixed
points (actually manifolds) rests on the special properties that we
have established for the renormalization of the superpotential
\eqref{uu1}
and for the gauge coupling \eqref{hh1} or \eqref{hh2}. 

First of all, consider a Wess-Zumino model (so only chiral
multiplets). In that case, one can prove that all the anomalous
dimensions $\gamma_i\geq0$ (with equality for a free theory). 
Since, in order to have an interacting theory, we need terms cubic in
the superpotential, it follows immediately that there cannot be
non-trivial fixed points of RG:
\EQ{
\mu\frac{d\lambda}{d\mu}=(-3+n+n\gamma)\lambda>0\ .
}

Now we add vector multiplets and take the chiral multiplets in
representations $r_f$ of the gauge group. In that case, there is no positivity
conditions on the anomalous dimensions of the chiral multiplets. In
order to have a fixed point, for each coupling in the superpotential
$\sim\lambda\phi_{f_1}\cdots\phi_{f_p}$,
we need from \eqref{uu1}
\EQ{
\sum_{i=1}^p\gamma_{f_i}=3-p
\label{k11}
}
and from \eqref{hh1}
\EQ{
3C(G)-\sum_f C(r_f)\Big(1-2\gamma_f\Big)=0\ .
\label{k12}
}
If there are $n$ such cubic couplings $\lambda_i$ 
in the superpotential then it
appears that there are $n+1$ equations 
for $n+1$ unknowns $\{\lambda_i,g\}$. Generically, therefore,
if solutions exist they will be discrete and unlikely to be within the
reach of perturbation theory. However, there are special situations when the
set of $n+1$ equations are degenerate in
which case there are spaces of fixed points which can extend into the
perturbative regime.

For example, suppose there are 3 chiral multiplets in the adjoint
representation with a superpotential which has sufficient symmetry to
infer that all the anomalous dimensions are equal
$\gamma_{f}\equiv\gamma$. In this case, 
\eqref{k11} and \eqref{k12} are satisfied if the superpotential is
cubic in the fields and if the anomalous dimension
\EQN{
\gamma(\lambda_i,g)=0 
}
which is a single condition 
on $n+1$ couplings. Such theories are very special because they
are actually {\it finite\/}. 

\BOX{\begin{center}{\bf Finite Theories}\end{center}

A theory is {\it finite\/} if there are no UV divergences in perturbation
theory. In a SUSY theory, this means that the anomalous dimensions of
all the chiral operators vanish and the beta function of the gauge
coupling vanishes. The conditions are:\\

(i) $\gamma_f=0$.\\

(ii) The superpotential must be cubic in the fields, in order that the
RG flow of the couplings in the superpotential vanish: see \eqref{uu1}.\\

(iii) $3C(G)=\sum_f C(r_f)$ in order that the RG flow of
the gauge coupling vanishes. \\

Notice that not every finite theory is a CFT since conformal
invariance could be broken by VEVs for scalar fields if there is a
moduli space of vacua. However, in such cases
conformal invariance would be recovered
in the UV. Neither is every conformal field theory a finite theory
since the condition to be at a fixed point does not require the
anomalous dimensions of fields to vanish.
}

For example, if the gauge group is $G=SU(N)$, there are 3 gauge
invariant couplings one can write in the superpotential for 3
adjoint-valued fields ($N\times N$ traceless Hermitian matrices) which have
enough symmetry to imply that all the anomalous dimensions are equal:
\EQ{
W=\text{tr}\,\big(\lambda_1\phi_1\phi_2\phi_3+\lambda_2\phi_1\phi_3\phi_2
+\frac{\lambda_3}3\sum_{f=1}^3\phi_f^3\big)\ .
\label{spw}
}
Of course we need to check that the condition
$\gamma(\lambda_1,\lambda_2,\lambda_3,g)=0$
actually has solutions. The key to proving this is to establish that
solutions exist in perturbation theory and then by continuity one can expect
that solutions exist also at strong coupling. To one-loop it can be
shown that the anomalous dimension is$^{\ref{t0}}$
\EQ{
\gamma=\frac{C(G)}{64\pi^2}\Big(\sum_f|\lambda_f|^2-4g^2\Big)\ .
\label{exg}
}
We won't give the proof of this, but it is easy to see the diagrams
that contribute. For example, for the anomalous dimension of the scalar
field, the following 1-loop graphs contribute (plus graphs involving the ghosts
which we do not show)
\begin{center}
\includegraphics[height=2in]{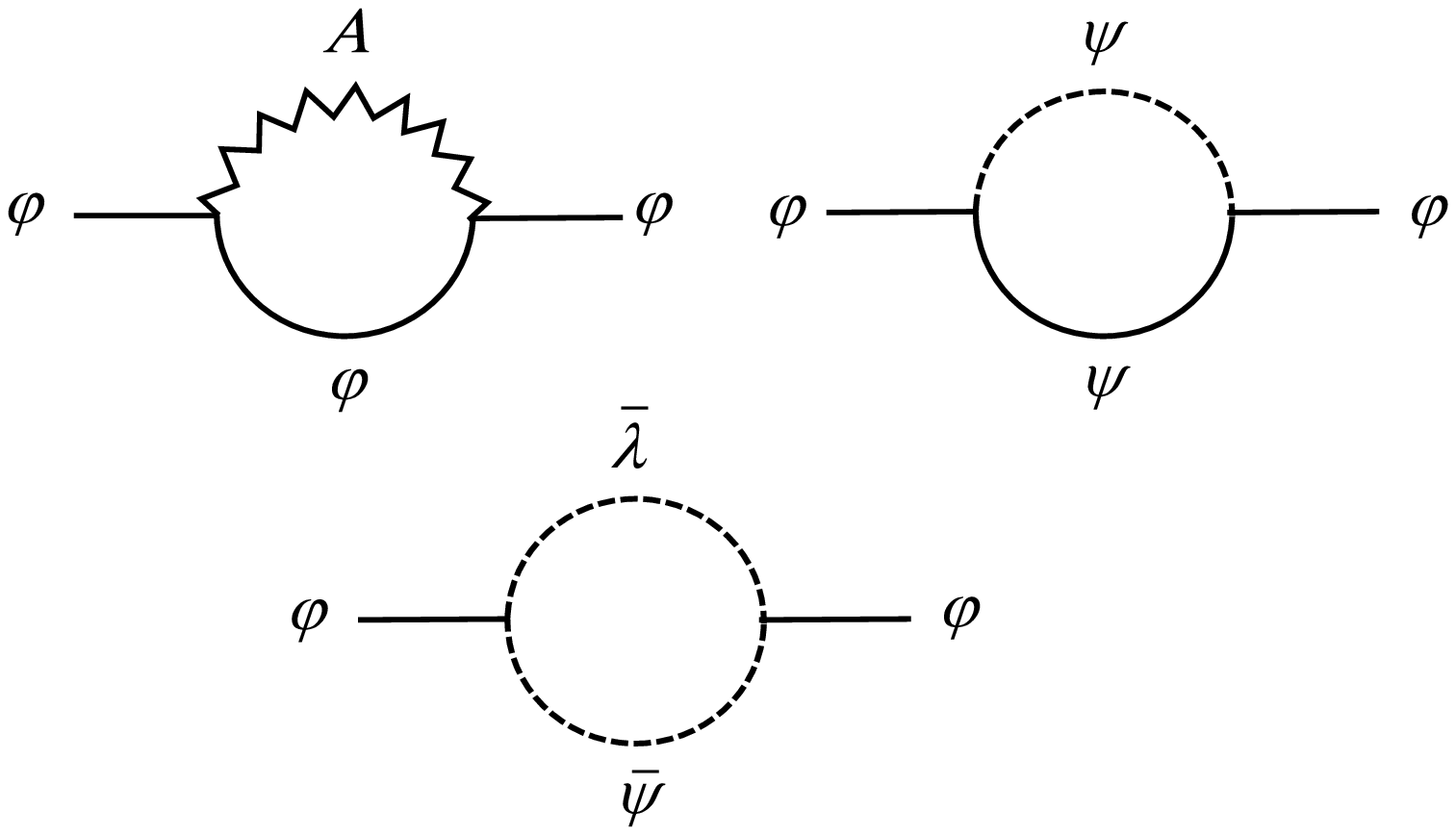}
\end{center}
So $\gamma$ receives positive contributions 
from the chiral multiplets and
negative contributions from the vector multiplet. To simplify the
discussion suppose that all the couplings $\lambda_f\sim\lambda$. 
In this case, there is a RG-fixed line at weak coupling when $\lambda\sim g$.
RG flow in the $(\lambda,g)$ subspace is of the form
\begin{center}
\includegraphics[height=2in]{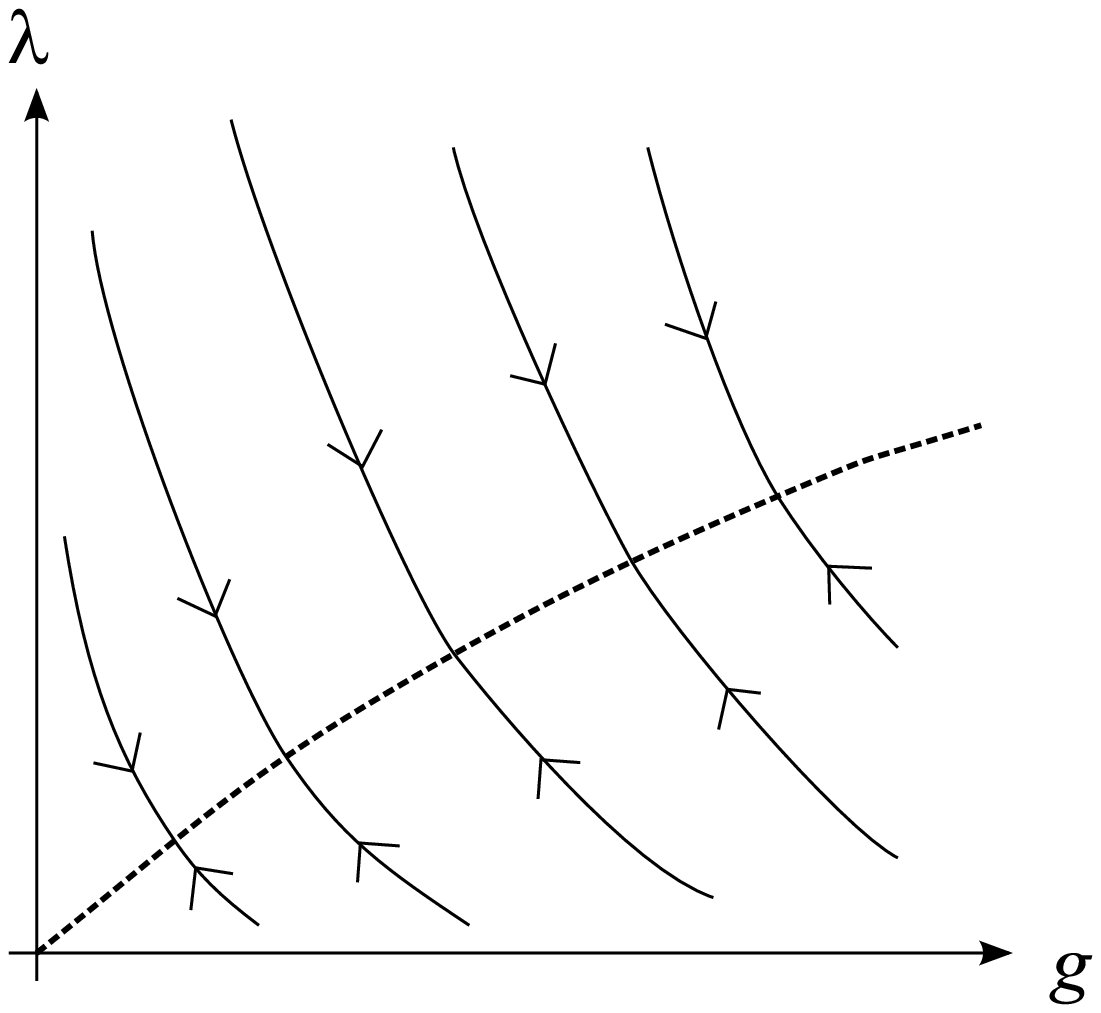}
\end{center}
where the dotted line is a line of fixed points. 
So the couplings away from the fixed line  
in the $(\lambda,g)$ subspace are irrelevant since the RG flow is
into the fixed line. One would expect that the fixed line extends by
continuity into the region of strong coupling as well.

Once we take all the couplings into account, there is actually a
6-dimensional space of SCFTs (this comes from the complex $\lambda_f$,
$f=1,2,3$ and the gauge coupling subject to one condition). 

\noindent{\bf\sl The maximally SUSY gauge theory}

A special class of these finite theories corresponds to taking a SUSY gauge
theory with 3 chiral multiplets in the adjoint representation
with a superpotential of the form \eqref{spw} with
$\lambda_1=-\lambda_2=\sqrt2g$ and $\lambda_3=0$, {\it i.e.\/}$^{\ref{t0}}$
\EQN{
W=\sqrt2g\text{Tr}\,\phi_1[\phi_2,\phi_3]\ .
}
Such theories have {\it extended\/} $\N=4$ SUSY, the 
maximal amount of the SUSY in $d=4$ for theories which do not 
contain gravity.

The theory has a large global $SU(4)$ symmetry under which the 3
fermions from the chiral multiplets and the gluino transform as the
${\bf 4}$, while the 3 complex scalars $\phi_i$ can be written as 6
real scalars that transform in the antisymmetric ${\bf 6}$ (or the
vector of $SO(6)\simeq SU(4)$). Such symmetries of SUSY theories for
which the fermions and scalars transform differently are known as
R-symmetries. It is no accident that $SO(6)$ is the isometry group of
an $S^5$: see below. For later use, the potential of the theory has
the form
\EQ{
V(\phi_i,\phi^*)=g^2\sum_{ij=1}^3\text{Tr}\big(\phi_i\phi_j\phi_i^*\phi_j^*-
\phi_i\phi_j\phi_j^*\phi_i^*\big)\ .
\label{pot}
}
$\N=4$ are infinitely fascinating due to the amazing
fact that they are equivalent to Type IIB string theory in 10-d
spacetime on an $AdS_5\times S^5$ background. The fact that what
appears to be a humble 4d gauge theory can encode all the rich
dynamics of a ten-dimensional (graviational) string theory is, I believe,  
one of the most amazing results in theoretical physics.
We won't give a complete discussion of the AdS/CFT correspondence here
since it will be discussed at length in other lectures at this
school. However, since RG plays an important
r\^ole, I will set the scene. 

The gauge theory has 2 parameters $g$ and $N$. In the dual
string picture the {\it string coupling\/} $g_s=g^2/4\pi$ while the
radius of the geometry in string units is $\sqrt{g^2N}$. This
motivates the 't~Hooft limit in which $N\to\infty$ with $\lambda=g^2N$
fixed. This is the so-called {\it planar\/} limit on the gauge theory
side since only planar Feynman graphs survive, while the string
side should correspond to free strings moving in an $AdS_5\times S^5$
geometry. Perturbation theory on the gauge theory side requires
$\lambda$ small which is the limit on the string side where the
geometry is highly curved. While strong coupling, 
large $\lambda$, in the gauge theory corresponds to strings moving on a
weakly curved background.

Part of the ``dictionary'' is that single trace composite 
operators of the form
\EQN{
{\cal O}(x)=\text{tr}\,\big(a_1a_2\cdots a_L\big)\ ,
}
where the $a_i$ are one of the fundamental fields (where $A_\mu$
appears through ${\cal D}_\mu$ in order to be gauge invariant)
correspond to single string states. Moreover the scaling dimension of
the operator equals the energy of the associated string state:
\EQN{
{{\mathscr E}}_\text{string}=\Delta_{\cal O}\ .
}
So perturbative calculations of the anomalous dimensions of operators
in the gauge theory tells us directly about the spectrum of strings
moving in a highly curved $AdS\times S^5$ geometry!

There have been some amazing developments in calculating these
anomalous dimensions in the gauge theory and matching them to the
energies of string states. Here, we will  
consider the problem of calculating the anomalous dimensions
of single trace operators made up of just the two ``letters'' $\phi_1$ and
$\phi_2$ to one-loop in planar perturbation theory (perturbation
theory in $\lambda$ with $N=\infty$ which suppresses
completely non-planar contributions):
\EQN{
{\cal O}=\text{tr}\,\big(\phi_1\phi_1\phi_2\cdots\big) 
}
If the operator has length $L$ then the classical dimension is simply
$d_{\cal O}=L$, so
\EQN{
\Delta_{\cal O}=L+\lambda\Delta_1+\lambda^2\Delta_2+\cdots\ .
}
The problem is that the operators ${\cal O}_i$ with a given number $J_1$ of
$\phi_1$ and $J_2$ of $\phi_2$, with $J_1+J_2=L$, all mix under RG and
so we have the problem of diagonalizing a matrix.

One way to calculate the anomalous dimension of the class of
such operators with fixed $J_1$ and 
$J_2$, $\big\{{\cal O}_p\big\}$, is to add them to the action
\EQN{
S\longrightarrow S+\int d^dx\,\sum_p \mu^{4-L}g_p{\cal O}_p(x)
}
and then look at the flow of the $g_p$ in the effective potential to
linear order in $g_p$. We follow exactly the same 
background field method that we used earlier and treat the operator terms
as new vertices in the action with couplings $g_p$. 
The flow of the couplings $g_p$ can be deduced by
writing down Feynman graphs with $J_1$ external $\phi_1$ and $J_2$
external $\phi_2$ lines with fluctuating fields on internal lines. 

Since the anomalous dimension follows from
\EQN{
\mu\frac{dg_p}{d\mu}=(L-4)g_p+\gamma_{pq}g_q+\cdots\ ,
}
we only need look at graphs which use the vertices $g_p$ once.  
At one-loop level, to begin with we have the graphs
\begin{center}
\includegraphics[width=3.5in]{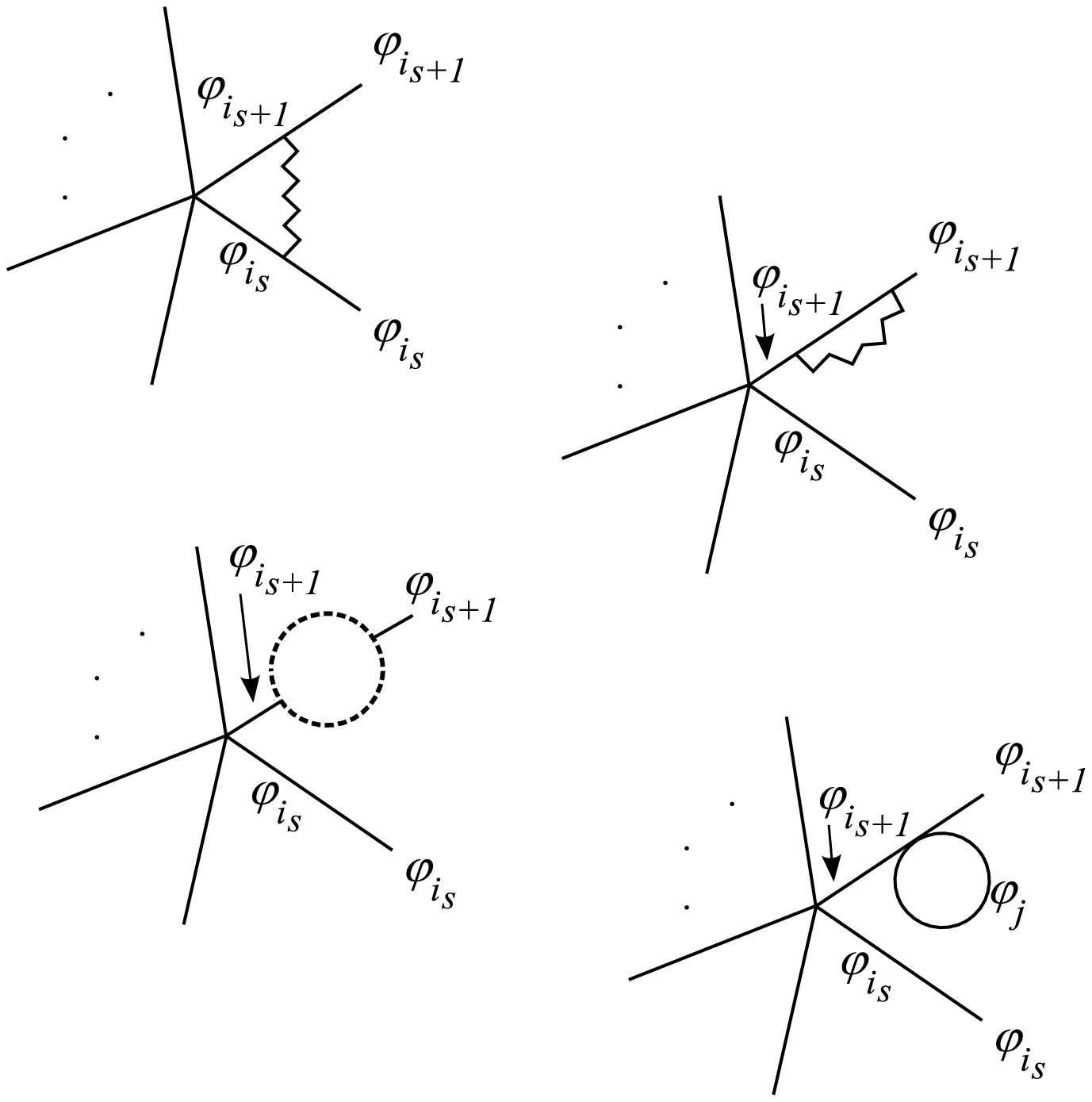}
\end{center}
The first graph has to be summed over all neighbouring pairs (but not
non-neighbouring pairs, since these would be non-planar graphs
suppressed by powers of $1/N$), while the others
are to summed 
over all $L$ legs. The third involves a
fermion loop and the fourth a scalar loop. The important point about these
graphs is the they don't change the flavour of the legs of the
vertex. In other words, whatever their contribution to the anomalous
dimension is proportional
to the identity in the space of $(J_1,J_2)$ operators. We shall shortly
argue that these contributions, which we write 
$C_1{\bf 1}$, actually vanish, so $C_1=0$.

The remaining graphs involve using the quartic coupling in the scalar
potential to tie two adjacent legs together. 
The potential \eqref{pot} contains the terms
\EQN{
V=2g^2\text{tr}\big(\phi_1\phi_2\phi_1^*\phi_2^*-\phi_1\phi_2\phi_2^*\phi_1^*
-\phi_2\phi_1\phi^*_1\phi_2^*+\phi_2\phi_1\phi_2^*\phi_1^*\big)+\cdots\ ,
}
and so we see immediately that these interactions can be used to form two
additional one-loop graphs when two adjacent legs are different,
either $\phi_1\phi_2$, as shown, or $\phi_2\phi_1$
\begin{center}
\includegraphics[width=3.5in]{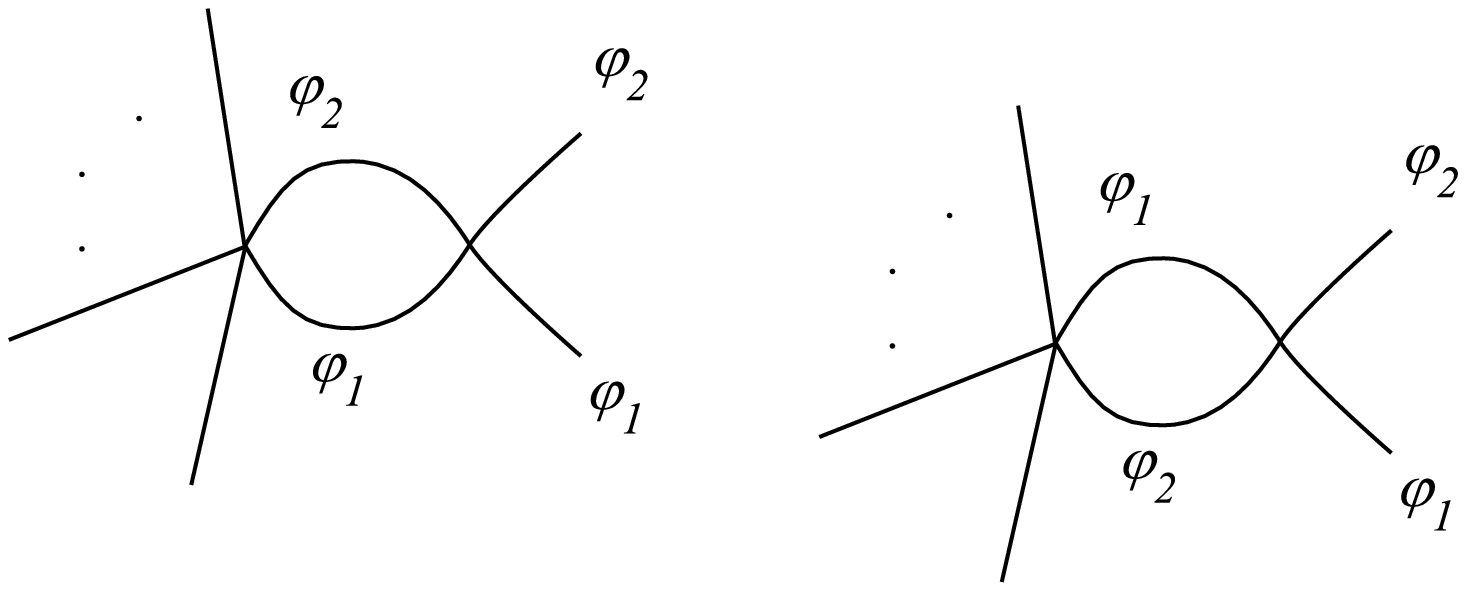}
\end{center}
The absolute
contribution is simple to calculate, but even without explicit
calculation it is easy to see
from the potential 
that these contributions come with a relative $-1$.

Putting all this together, we have found that the two sets of one-loop
contributions to the
anomalous dimension ``operator'' can be written neatly as
\EQN{
\gamma=\sum_{\ell=1}^L\Big\{C_1{\boldsymbol 1}_{\ell}+C_2
\big({\boldsymbol 1}_{\ell}-{\boldsymbol P}_{\ell}\big)\Big\}
}
where ${\boldsymbol P}_{\ell}$ permutes the $\ell^\text{th}$ and
$\ell+1^\text{th}$ letters in the word and ${\bf 1}_\ell$ is the identity:
\EQN{
{\boldsymbol P}_\ell\,\,\text{tr}\big(\cdots\phi_{i_\ell}\phi_{i_{\ell+1}}
\cdots\big)=\text{tr}\big(\cdots\phi_{i_\ell+1}\phi_{i_{\ell}}
\cdots\big)
}
and we identify $L+1\equiv1$ due to the cyclicity of the trace. 

Now we can pin down $C_1$ by using the fact that 
the operator $\text{tr}\,\phi_1^L$ is special because it is a
so-called BPS operator and is consequently 
protected against quantum corrections; hence, $\Delta=L$ to all orders
in the perturbative expansion. This means that $C_1=0$ a simple
computation gives
\EQN{
C_2=\frac{\lambda}{4\pi^2}\ .
}

Rather interestingly the resulting anomalous dimension operator
$\gamma$ up to some overall scaling is identical to the Hamiltonian of
the so-called $XXX$ spin chain, a quantum mechanical model of $L$
spins taking values $|\uparrow\rangle\equiv\phi_1$ and
$|\downarrow\rangle\equiv\phi_2$. Each operator of fixed length 
corresponds to a state of the spin chain:
\EQN{
\text{tr}\big(\phi_1\phi_1\phi_2\phi_1\phi_2\phi_2\phi_1\big)
\longleftrightarrow
|\uparrow\uparrow\downarrow\uparrow\downarrow\downarrow\uparrow
\rangle\ .
}
The problem of finding the anomalous
dimensions and hence the spectrum of string states then is identical
to the problem of finding the eigenstates of the spin chain, a
problem that was solved by Bethe in 1931 by means of what we now call
the {\it Bethe Ansatz\/} which reflects the fact that the problem is
in the special class of integrable system. 
This observation is just the beginning of
the fascinating story of $\N=4$, integrability and the AdS/CFT
correspondence.

\begin{center}{\bf Notes}\end{center}

\begin{quote}
{\small

\begin{note}The coupling here and in the following is generally the
  canonical gauge coupling, however, we won't distinguish the $g$ and
  $g_c$ for now on.
\label{t0}
\end{note}

\begin{note}$F$ is also complex, while $D$ is Hermitian and for the
  fermions $\psi^\dagger_\alpha=\bar\psi^\aD$ and 
$\lambda^\dagger_\alpha=\bar\lambda^\aD$.
\label{t1}
\end{note}

\begin{note}The perturbative expansion is really an expansion in
  $g^2N$. In our limit, however, $g^2\sim1/N^2$ and so higher orders
  in perturbation theory are suppressed by $1/N$.
\label{s10}
\end{note}

}\end{quote}

\end{document}